\documentclass[11pt]{article}
\usepackage{epsfig}
\newcommand{\la}[1]{\label{#1}}
\newcommand{\be}{\begin{equation}}
\newcommand{\ee}{\end{equation}}
\newcommand{\ba}{\begin{eqnarray}}
\newcommand{\ea}{\end{eqnarray}}
\newcommand{\bi}{\begin{itemize}}
\newcommand{\ei}{\end{itemize}}
\newcommand{\rmi}[1]{{\mbox{\scriptsize #1}}}
\newcommand{\nr}[1]{(\ref{#1})}
\newcommand{\tr}{{\rm Tr\,}}
\newcommand{\str}{{\rm Str\,}}
\newcommand{\re}{\mathop{\rm Re}}

\newcommand{\nn}{\nonumber \\}
\newcommand{\fr}[2]{{\frac{#1}{#2}}}

\renewcommand{\vec}[1]{{\bf #1}}

\newcommand{\Nf}{N_{\rm f}}
\newcommand{\Nv}{N_{\rm v}}
\newcommand{\Pv}{P_{\rm v}}
\newcommand{\rmO}{{\rm O}}
\newcommand{\RR}{{\rm I\kern -.2em  R}}
\newcommand{\eq}{Eq.~}
\newcommand{\eqs}{Eqs.~}
\newcommand{\fig}{Fig.~}

\newcommand{\se}{Sec.~}

\def\lsi{\raise0.3ex\hbox{$<$\kern-0.75em\raise-1.1ex\hbox{$\sim$}}}
\def\gsi{\raise0.3ex\hbox{$>$\kern-0.75em\raise-1.1ex\hbox{$\sim$}}}
\newcommand{\lsim}{\mathop{\lsi}}
\newcommand{\gsim}{\mathop{\gsi}}

\makeatletter \@addtoreset{equation}{section} \makeatother
\renewcommand{\theequation}{\arabic{section}.\arabic{equation}}
\makeatletter
\renewcommand\section{\@startsection {section}{1}{\z@}%
                                   {-5.5ex \@plus -1ex \@minus -.2ex}
                                   {2.3ex \@plus.2ex}%
                                   {\normalfont\large\bfseries}}
\renewcommand\subsection{\@startsection{subsection}{2}{\z@}%
                                     {-3.25ex\@plus -1ex \@minus -.2ex}%
                                     {1.5ex \@plus .2ex}%
                                     {\normalfont\normalsize\bfseries}}
\renewcommand\thesection {\@arabic\c@section}
\renewcommand\thesubsection   {\thesection.\@arabic\c@subsection}
\renewcommand{\@seccntformat}[1]{%
\csname the#1\endcsname.\hspace{1.0em}}
\makeatother

\begin{document}

\begin{titlepage}
\begin{flushright}
\begin{minipage}[t]{6.2cm}
BI-TP 2003/35     \hfill FTUV-03-1208      \\
CERN-TH/2003-293  \hfill IFIC/03-54       \\
CPT-2003/PE.4617  \hfill MPP-2003-129      \\
DESY-03-195       \hfill hep-lat/0312012
\end{minipage}
\end{flushright}

\begin{centering}

\vspace*{0.8cm}

{\Large\bf 
Low-energy couplings of QCD from topological} 

\vspace*{0.3cm}

{\Large\bf 
zero-mode wave functions}

\vspace*{0.8cm}

L.~Giusti$^{\rm a,}$\footnote{leonardo.giusti@cern.ch}$^{\rm ,}$%
\footnote{Address since Dec 1, 2003: 
CPT, 
CNRS, Case 907, 
Luminy, F-13288 Marseille, France.}, 
P.~Hern\'andez$^{\rm b,}$\footnote{pilar.hernandez@ific.uv.es}, 
M.~Laine$^{\rm c,}$\footnote{laine@physik.uni-bielefeld.de}, 
P.~Weisz$^{\rm d,}$\footnote{pew@mppmu.mpg.de}, 
H.~Wittig$^{\rm e,}$\footnote{wittig@mail.desy.de}

\vspace*{0.8cm}

{\em $^{\rm a}$%
Theory Division, CERN, CH-1211 Geneva 23,
Switzerland\\}
 
\vspace{0.3cm}

{\em $^{\rm b}$%
Depto.\ de F\'isica Te\'orica and IFIC, 
Universidad de Val\`encia, 
E-46100 Burjassot, Spain\\}
\vspace{0.3cm}

{\em $^{\rm c}$%
Faculty of Physics, University of Bielefeld, 
D-33501 Bielefeld, Germany\\}

\vspace{0.3cm}

{\em $^{\rm d}$%
Max-Planck-Institut f\"ur Physik, F\"ohringer Ring 6, 
D-80805 Munich, Germany\\}

\vspace{0.3cm}

{\em $^{\rm e}$%
DESY, Theory Group, Notkestrasse 85, 
D-22603 Hamburg, Germany\\}

\vspace*{0.8cm}
  
\mbox{\bf Abstract}

\end{centering}

\vspace*{0.3cm}

\noindent
By matching $1/m^2$ divergences in finite-volume two-point 
correlation functions of the scalar or pseudoscalar densities with 
those obtained in chiral perturbation theory, we derive a 
relation between the Dirac operator 
zero-mode eigenfunctions at fixed non-trivial topology 
and the low-energy constants of QCD. 
We investigate the feasibility of using this relation
to extract the pion decay constant, by computing  
the zero-mode correlation functions on the lattice in the 
quenched approximation and comparing them with the 
corresponding expressions in quenched chiral 
perturbation theory.
\vfill

 
\noindent
January 2004

\vfill

\end{titlepage}
\setcounter{footnote}{0}

%
\section{Introduction}

It is well appreciated that a combination of lattice methods and chiral 
perturbation theory ($\chi$PT) can be an efficient tool for studying the 
low-energy properties of QCD close to the chiral limit. While $\chi$PT 
is the perfect book-keeping device for the non-trivial relations implied 
by chiral symmetry, the lattice can be used to determine the low-energy 
couplings of this theory, which encode the dynamics of 
the fundamental Lagrangian.

The study of QCD on the lattice obviously requires a finite volume, and this 
might appear problematic close to the chiral limit, since spontaneous 
chiral symmetry breaking does not take place in a finite volume. 
This is not the case, however, because $\chi$PT is 
able to predict analytically the large finite-size effects expected 
in this regime, in terms of the same low-energy constants as appear 
in an infinite volume, in such a way that infinite-volume quantities 
can be obtained unambiguously from the finite-volume ones. 
The study of $\chi$PT in a finite volume and close to the chiral limit 
(in the so-called $\epsilon$-regime) was pioneered by Gasser and 
Leutwyler~\cite{GL}--\cite{h} a long time ago, but it is only 
recently that practical ``measurements'' of physical observables 
became feasible in lattice QCD~\cite{condensate}--\cite{glww}. 
This is thanks to the new formulations of lattice fermions, which 
preserve an exact chiral symmetry~\cite{gw}--\cite{kn}.
In this paper we will employ one of these formulations and invoke the  
specialized numerical techniques developed in ref.~\cite{methods}, 
which are needed for high-precision studies 
in the $\epsilon$-regime.

It was found in \cite{ls} that in the $\epsilon$-regime,  
gauge field topology may play a very important role.
In a given chiral regularization of QCD, averages  can be defined
in sectors of fixed topological index $\nu$~\cite{hln}, and our
assumption will be that standard ultraviolet renormalization 
also makes sense in such sectors. Although this is a non-trivial 
assumption in QCD, there is a well-defined prescription for
how to compute analogous averages in $\chi$PT.

It then turns out that close to the chiral limit, 
many observables depend quite strongly on the topology.
In particular, for $\nu\neq 0$, two-point functions of the scalar 
and pseudoscalar densities have poles in the quark mass squared,
with residues given by correlation functions of Dirac operator 
zero-mode eigenfunctions. In the $\epsilon$-regime of $\chi$PT
the same poles appear, with residues that are calculable 
functions of the low-energy constants. Requiring the residues in 
the fundamental and effective theories to be the same yields 
non-trivial relations.

To be more specific, at leading order in $\chi$PT the correlators
mentioned are constants depending only on $\nu$ and the volume, 
but at next-to-leading order (NLO) one obtains a 
space-time-dependent function, which also  
involves the pseudoscalar decay constant $F$. 
Therefore, $F$ can be determined by monitoring the amplitude 
of the time dependence.
A nice feature of this procedure is that it does not require 
knowledge of renormalization factors since we employ 
a regularization that preserves the chiral symmetry.

Given that $F^2$ appears first at the NLO, $\rmO(1/F^2)$, and that
the convergence of $\chi$PT at realistic (not very large) volumes is not 
{\it a priori} guaranteed to be rapid, it is one of the purposes of this paper 
to present the results of the calculation up to the next-to-next-to-leading 
order (NNLO), $\rmO(1/F^4)$. According to our conventions 
as detailed in~\ref{se:largeNc}, these correspond
to the relative orders $\rmO(\epsilon^4), \rmO(\epsilon^8)$
in the $\epsilon$-expansion, respectively. 

We study, furthermore, the feasibility of using this relation 
to extract $F^2$ from the zero-mode wave functions 
computed on the lattice, in the quenched approximation. 
Thus, predictions for the quenched version of $\chi$PT (Q$\chi$PT)
(whose theoretical status is unfortunately rather questionable, see 
\se\ref{cqxpt} and, e.g., ref.~\cite{pd}) 
are also presented, at the same order. 

The paper is organized as follows. In \se\ref{sec:full} we derive the
relation alluded to above and present the 
results of the calculation of the pseudoscalar density correlator 
in full $\chi$PT. In \se\ref{sec:feasibility} we obtain the same 
results in the quenched approximation and compare them with a numerical 
determination of the zero-mode eigenfunctions in lattice QCD, using 
overlap fermions. We conclude in \se\ref{sec:conclu}, and 
collect various details of the NNLO computations in 
three Appendices.

%
\section{Pseudoscalar correlator in QCD and in $\chi$PT}
\la{sec:full}

\subsection{The fundamental theory}

In this paper we are concerned with QCD in a finite volume $V=T\times 
L^3$, with periodic boundary conditions in all directions.
Our conventions for the Dirac matrices are such that 
$\gamma_\mu^\dagger = \gamma_\mu$, 
$\{\gamma_\mu,\gamma_\nu\} = 2 \delta_{\mu\nu}$, 
$\gamma_5 = \gamma_0 \gamma_1 \gamma_2 \gamma_3$, 
so that the (unquenched) Euclidean continuum 
quark Lagrangian formally reads
\be
  \mathcal{L}_E = \bar \psi (\gamma_\mu D_\mu + M )\psi
  \;, 
\ee
where $M$ is the mass matrix. For simplicity, we take
$M$ to be diagonal and degenerate, $M = \mathop{\mbox{diag}}(m,...,m)$. 
The number of dynamical flavours appearing in $\psi$ is 
denoted by $\Nf$.

In the following we will restrict our attention to correlation functions 
of the scalar and pseudoscalar densities,
\ba
 \mathcal{S}^I  \equiv  \bar\psi T^I \psi\;,  
 \;\;\;\;\;\; \mathcal{P}^I  \equiv  \bar\psi i \gamma_5 T^I \psi
 \;, \la{SP} 
\ea
involving $\Nv$ valence quarks; in the unquenched theory, $\Nv \equiv 
\Nf$. The $\Nv\times\Nv$ valence flavour basis is generated by 
\be
 T^I \equiv \{ T^0, T^a \}, \quad
 T^0 \equiv I_{\Nv}, \quad a = 1,...,\Nv^2-1
 \;, \la{TI}
\ee
where ${I_{\Nv}}$
is the $\Nv\times\Nv$ identity matrix,  and
the traceless $T^a$ are assumed to be normalized so that 
\be
\tr [T^a T^b] = \fr12 \delta^{ab}  \;. 
\la{Tnorm}
\ee

Our analysis is based on the assumption that 
correlation functions at fixed topology, e.g. the two-point correlators
of pseudoscalar densities,
\be
\mathcal{C}^{IJ}_\nu(x-y)=\Bigl\langle
 \mathcal{P}^I(x) \mathcal{P}^J(y) \Bigr\rangle_\nu \;,
\la{cijdef}
\ee
have a well-defined meaning in the continuum limit at non-zero
physical distances.
Although plausible, this is a non-trivial 
dynamical issue and to pose precise questions we must 
introduce an ultraviolet regularization. 

We here adopt the lattice regularization with a massless 
Dirac operator $D$ obeying the Ginsparg--Wilson (GW) relation,
since it preserves an exact chiral symmetry. The topological index
assigned to a configuration then is $\nu=n_+-n_-$,
where $n_+$ ($n_-$) are the numbers of zero-modes of $D$ with 
positive (negative) chirality. Correlation functions such as 
\eq(\ref{cijdef}) are now well defined at fixed cutoff~\cite{hln}, 
and the question is whether, in any given sector of index $\nu$,
they have a continuum limit independent of the particular
choice of $D$ %
  \footnote{Since the space of lattice gauge fields is connected, 
  different choices of $D$ possibly lead to different assignments
  of index for a given configuration.}.
Our working hypothesis is that this is indeed the case;
some recent numerical evidence (in the quenched approximation) 
consistent with this scenario can be found, e.g. in 
refs.~\cite{glww,ddp}.

By employing the spectral representation of the quark propagator, 
it is clear that the correlator in \eq(\ref{cijdef}) contains 
a pole in $m^2$, due to the exact zero modes. Its residue is
\be
 \lim_{m\rightarrow 0} (m V)^2~\mathcal{C}^{IJ}_\nu(x)
 =  \tr[T^I T^J]~\mathcal{A}_\nu(x) 
 + \tr[T^I] \tr[T^J]~{\tilde\mathcal{A}}_\nu(x)
 \;, 
\la{eq:spec} 
\ee
where 
\ba
 \mathcal{A}_\nu(x-y) & \equiv & \Bigl\langle 
 \sum_{i,j \in {\cal K}} v^\dagger_j(x) v_i(x) v^\dagger_i(y) v_j(y) 
 \Bigr\rangle_\nu 
 \;, \la{eq:v}
 \\ 
 {\tilde\mathcal{A}}_\nu(x-y) &\equiv & - ~ \Bigl\langle 
 \sum_{i \in {\cal K}} v^\dagger_i(x) v_i(x) 
 \sum_{j \in {\cal K}} v^\dagger_j(y) v_j(y) 
 \Bigr\rangle_\nu 
 \;, \la{eq:vbar}
\ea
and the sums are over the set 
of $|\nu|$ zero modes $v_i$ of the Dirac operator,
$D v_i =0 \; \forall \; i \in {\cal K}$, which
have definite chirality and are assumed to be normalized 
so that $\int \! {\rm d}^4 x \, v^\dagger_i(x) v_i(x) = V$.
\eq\nr{eq:v} corresponds to a ``connected'' contraction 
of the quark lines, \eq\nr{eq:vbar} to a ``disconnected'' one %
  \footnote{The terms in \eqs\nr{eq:v}, \nr{eq:vbar} 
  could also be interpreted as classical scattering amplitudes
  for pairs of zero modes.}.

It is important to note that in writing 
Eqs.~(\ref{eq:spec})--(\ref{eq:vbar})
we have assumed that poles arise only from exact zero modes, 
i.e. that taking the limit $m\to0$ and performing the 
average over the full space of configurations commute. 
At fixed volume the only potential danger 
arises from the average distribution of eigenvalues near zero;
our assumption holds if the density of eigenvalues
vanishes at fixed non-zero index. Intuitively one 
expects that distributions of non-zero eigenvalues at non-trivial topology 
are depleted near zero. In 
$\chi$PT, as well as in random matrix theory 
(\cite{phd} and references therein), 
the densities behave as $\rho_\nu(\lambda)\sim\lambda^{(2|\nu|+1+2 N_f)}$,
and no contribution from the non-zero modes is thus expected
in the observables we consider.

Since the zero modes are
eigenfunctions of $\gamma_5$, the scalar
and the pseudoscalar correlators contain the same information, 
\be
 \lim_{m\rightarrow 0} (m V)^2~\Bigl\langle 
\mathcal{S}^I(x)\mathcal{S}^J(y)\Bigr\rangle_\nu=
-\lim_{m\rightarrow 0} (m V)^2 {\cal C}^{IJ}_\nu(x-y)\,,
\ee
and hence we only consider the latter in the following. 
Finally we note that as a consequence of the exact chiral 
symmetry maintained by the GW lattice regularization,
the mass does not require additive renormalization and 
the products $m\mathcal{P}^I$ need no renormalization at all.

\subsection{Chiral perturbation theory}

At large distances, the two-point correlator of the
pseudoscalar density can be described by chiral perturbation theory.
The leading order chiral Lagrangian reads
\ba
 \mathcal{L}_\rmi{$\chi$PT} \!\! & = & \!\! \frac{F^2}{4} \tr 
 \Bigl[ \partial_\mu U \partial_\mu U^{\dagger} \Bigr] 
 - {m \Sigma \over 2} \tr
 \! \Bigl[ e^{i\theta/\Nf} U + U^{\dagger} e^{-i\theta/\Nf}\Bigr] 
 \;,
 \la{XPT}
\ea
where $U \in $ SU($\Nf$), and $\theta$ is the vacuum angle.  
This Lagrangian contains only two parameters, 
the pseudoscalar decay constant $F_{}$ and the 
chiral condensate $\Sigma$, while none of 
the higher order $L_i$ coefficients of Gasser and Leutwyler appear
at the next-to-leading non-trivial order
in the $\epsilon$-regime, $m\Sigma V\lsim 1$. 
The chiral theory
operator corresponding to $\mathcal{P}^I$ in \eq\nr{SP} reads, 
at leading order,
\ba
 {P}^I & = & 
 \;\; i \frac{\Sigma}{2} \tr \Bigl[ T^I 
  \Bigl( e^{i\theta/\Nf} U - U^{\dagger} e^{-i\theta/\Nf} \Bigr)\Bigr]
 \;.
\ea  
The correlators computed in $\chi$PT are referred to with the notation
\ba
 C^{II}_\nu (x-y) & \equiv & 
 \Bigl\langle P^I (x) P^I(y) \Bigr\rangle_\nu 
 \;, 
 \la{correlators}
\ea
where $I$ is not summed over, 
and the expectation value is taken at the topological index $\nu$. 

The correlators $C^{II}_\nu(x)$ have been computed by Hansen
in the $\epsilon$-regime without fixing the
topology~\cite{h}, 
up to relative order $\rmO(\epsilon^8)$,
according to our conventions for the counting rules of 
the $\epsilon$-expansion as they are specified in~\ref{se:largeNc}.
Our goal in this section is to repeat this calculation but {\it at fixed
topology}. 

Following the notation of \cite{h}, 
the general structure of the correlator is
(before volume averaging),
\be
 C^{II}_\nu(x) = 
  C_I + \alpha_I G(x) + \beta_I
 \Bigl[ G(x) \Bigr]^2 + 
 \gamma_I \int \! {\rm d}^4 y \,
 G(x-y) G(y) + 
 \epsilon_I \delta^{(4)}(x) 
  \;,
 \la{Cx}
\ee
where
\be
 G(x) = \frac{1}{V} 
 \sum_{n \in \mbox{\raisebox{-0.2ex}
 {$\scriptstyle\mathsf{Z\hspace*{-1.3mm}Z}^4$}}} 
 \Bigl(1 - \delta^{(4)}_{n,0} \Bigr) \frac{e^{i p \cdot x}}{p^2} 
 \;, \quad
 p = 2\pi\Bigl( \frac{n_0}{T}, \frac{\vec{n}}{L} \Bigr)
 \;.
 \la{Gx}
\ee
In dimensional regularization,
$G(0) = - \beta_1/\sqrt{V}$, with $\beta_1$ 
a dimensionless numerical coefficient
depending on the geometry of the box. 
According to~\eq\nr{Cx}, the result factorizes to terms 
representing space-time dependence, and to the coefficients
$C_I, \alpha_I, \beta_I, \gamma_I, \epsilon_I$, 
which turn out to contain integrals 
over the zero-mode Goldstone manifold. 
While all these quantities depend on the leading-order 
low-energy couplings $F$ and $\Sigma$, the constant $C_I$ and 
the contact term $\epsilon_I$ also 
depend on a combination of the $L_i$ coefficients
of Gasser and Leutwyler at the NNLO 
at which we are working~\cite{h}. 
To avoid the dependence on these
additional couplings we will only consider
the time variation of the correlators at non-zero 
time separations. 
For convenience, we also average the
correlators over the spatial volume $L^3$.

The various time dependences remaining after integration 
over the spatial volume are listed in Appendix~\ref{app:ints}, 
the emerging zero-momentum mode integrals 
in Appendix~\ref{app:zeromode}, and the expressions for the 
coefficients $C_I, \alpha_I, \beta_I, \gamma_I$ in terms of 
the zero-mode integrals in Appendix~\ref{app:coeffs}.
For $C_I$ the expressions are at NLO only, 
for the aforementioned reason.

Taking the volume average and considering the time derivatives 
of the residues of the $1/m^2$ poles, we define
\ba
 \lim_{m\to 0} (m V)^2 \frac{{\rm d}}{{\rm d}t}
 \int\! {\rm d}^3 \vec{x} \, C^{aa}_\nu (x) 
 & \equiv & {1\over 2} A'(t) 
 \;, \la{defA} \\
 \lim_{m\to 0} (m V)^2 \frac{{\rm d}}{{\rm d}t}
 \int\! {\rm d}^3 \vec{x} \, C^{00}_\nu (x) 
 & \equiv &  {\Nv}\, A'(t) + 
        {\Nv^2}\, \tilde A'(t) 
 \;. \la{deftA}
\ea
The spectral representation of \eq(\ref{eq:spec}) 
and the definitions $\mathcal{A}(t) = \int\!{\rm d}^3 \vec{x}\, 
\mathcal{A}_\nu(x)$,
$\tilde \mathcal{A}(t) = \int\!{\rm d}^3 \vec{x}\, \tilde \mathcal{A}_\nu(x)$, 
then imply that at large $t$,
\be
 \mathcal{A}'(t) = A'(t), \quad
 \tilde\mathcal{A}'(t) = \tilde A'(t) 
 \;. \la{relation}
\ee 
These constitute our basic relations between the zero-mode amplitudes 
and the pion decay constant in the chiral limit, $F$,
once we spell out the right-hand sides.
The latter actually 
vanish at the lowest order where
the undifferentiated quantities $A(t)=|\nu|L^3, 
\tilde{A}(t)=-\nu^2L^3$ are constant, 
matching the sum over volume of the zero-mode expressions
in \eqs(\ref{eq:v}), (\ref{eq:vbar}) \footnote{%
  Provided that the probability
  of having zero modes of both chiralities is zero.}.
On the other hand, given the expressions in \ref{app:xpt},  
we obtain, at NNLO, 
\ba 
 F^2 A'(t) & = &  \phantom{-}
 \frac{2 |\nu|}{\Nf}\biggl\{
(1 + \Nf |\nu|)h_1'(\tau)+\frac{T^2}{\Nf F^2 V}H_2(\tau) \biggr\}
 \;,
 \la{nonsinglet} \\
 F^2 \tilde A'(t) & = & 
 -\frac{2 |\nu|}{\Nf}
 \biggl\{
 (\Nf + |\nu|)h_1'(\tau)+\frac{T^2}{\Nf F^2 V}\tilde{H}_2(\tau) \biggr\} 
 \;,
 \la{singlet} 
\ea
where $\tau=t/T$, and the functions appearing are given by
\ba 
 H_2(\tau) & = &  
 -(1 + \Nf |\nu| )\Nf^2 \frac{\beta_1 \sqrt{V}}{T^2}h_1'(\tau) + 
 \Bigl[ 
 \Nf (6 - \Nf^2) |\nu| + 4 + \Nf^2 (2\nu^2-1)
 \Bigr] h_2'(\tau) \nn
 & & + 
 \Bigl[
 \Nf(2 - \fr12 \Nf^2)|\nu| + 1 + \fr12 \Nf^2 
 \Bigr] g_1'(\tau)
 \;,
 \la{nonsingleta} \\
 \tilde H_2(\tau) & = & 
 -(\Nf + |\nu|)\Nf^2 \frac{\beta_1 \sqrt{V}}{T^2}h_1'(\tau) +
 \Bigl[ (4 + \Nf^2) |\nu| + 2 \Nf(2 + \nu^2) - \Nf^3  \Bigr] h_2'(\tau) \nn 
 & & +  
 \Bigl[ (1 + \fr12 \Nf^2) |\nu| + 2\Nf - \fr12 \Nf^3 \Bigr] g_1'(\tau)
 \;.
 \la{singleta} 
\ea
The functions $h_1, h_2, g_1$ are defined in 
Eqs.~(\ref{h1})--(\ref{g1}).
Note that only the low-energy coupling $F$ appears here. 
A non-trivial check of these formulae is that, for $|\nu|=1$, they  
satisfy $A'(t)+ \tilde A'(t) = 0$ for any $\Nf$, 
as must be the case
since the sums in \eqs(\ref{eq:v}) and (\ref{eq:vbar}) 
are identical if there is only one zero mode.

\begin{figure}[t]



\epsfig{file=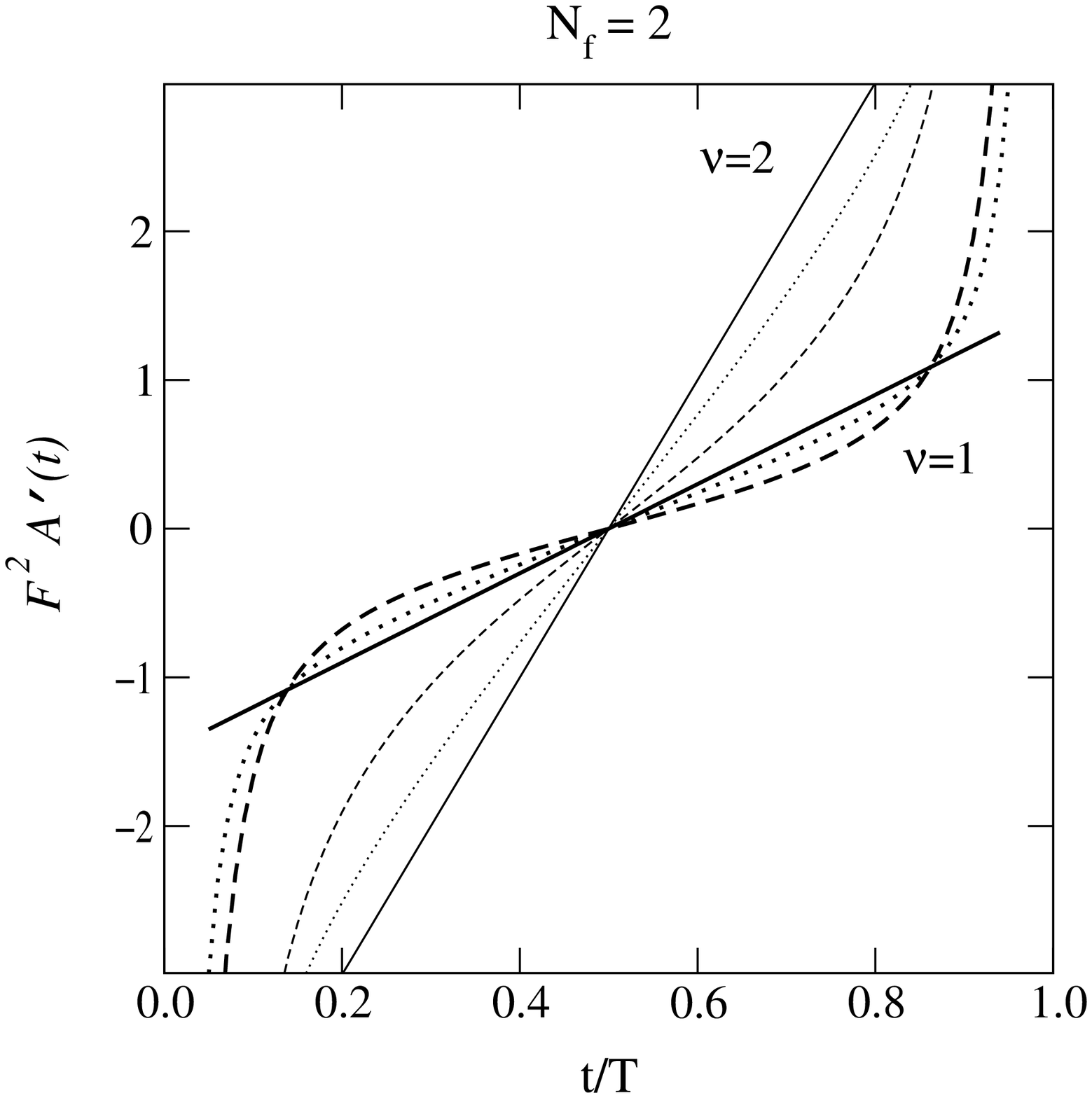,width=7.7cm}%
\epsfig{file=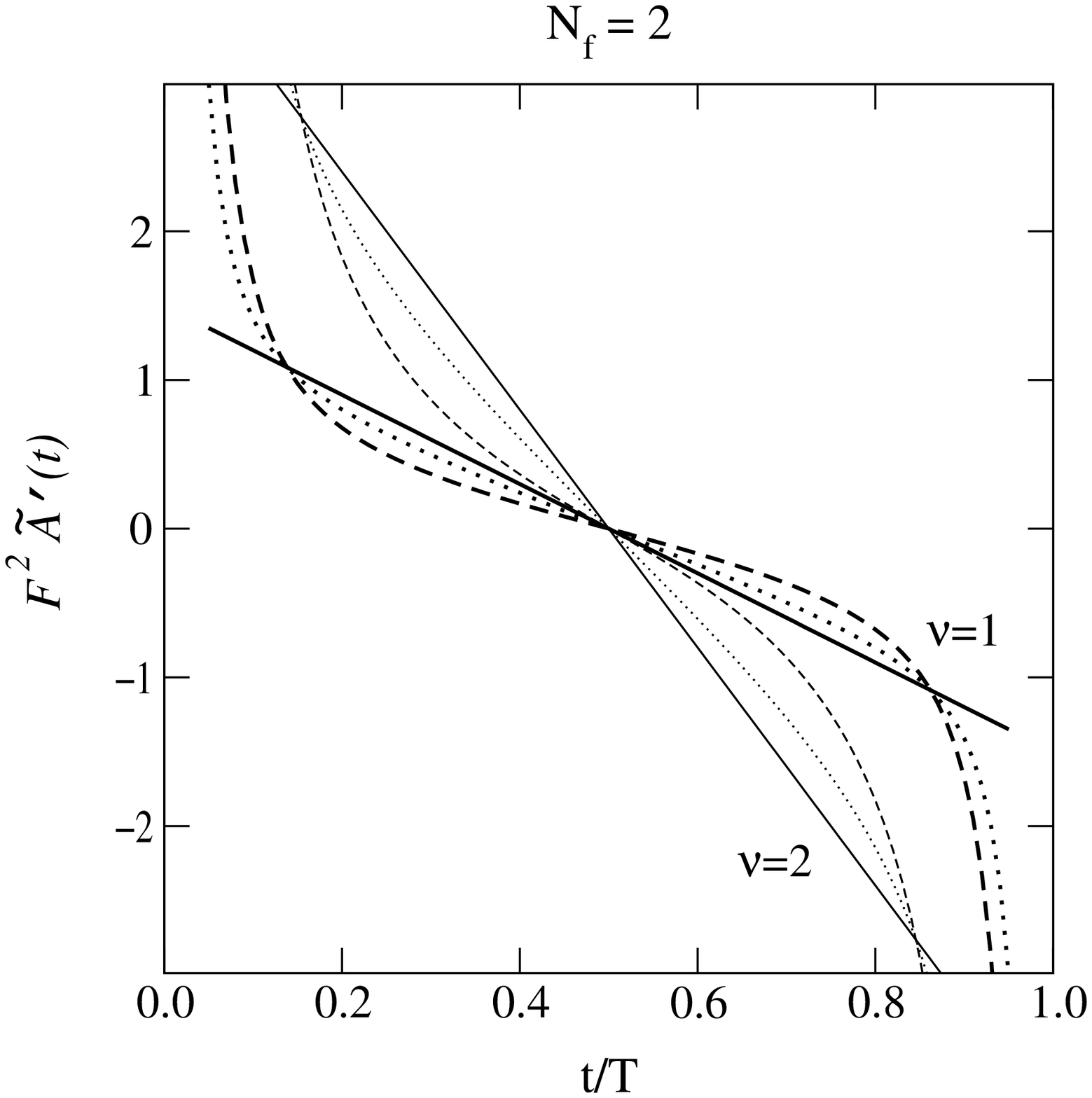,width=7.7cm}%

\vspace*{0.5cm}

\epsfig{file=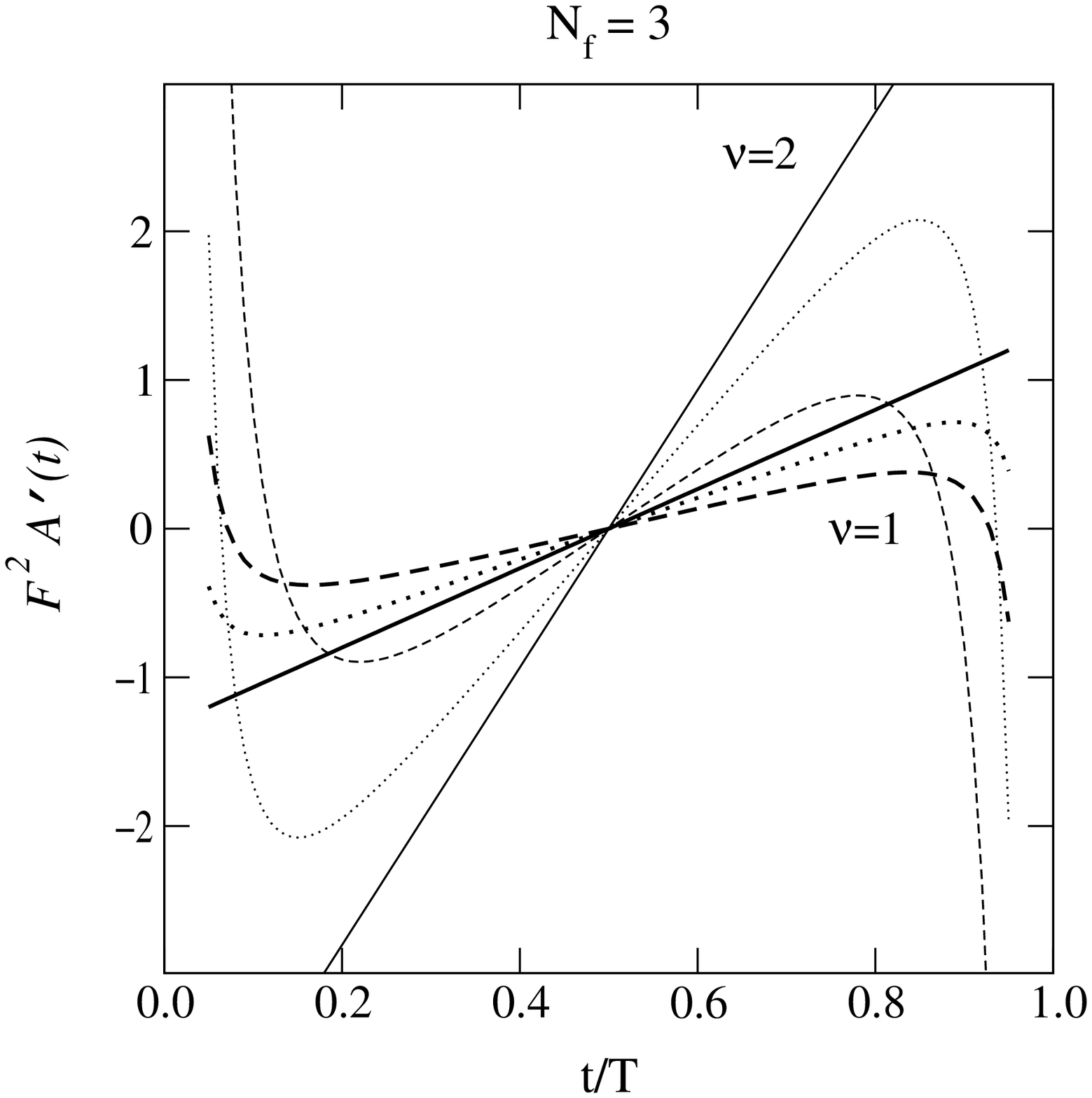,width=7.7cm}%
\epsfig{file=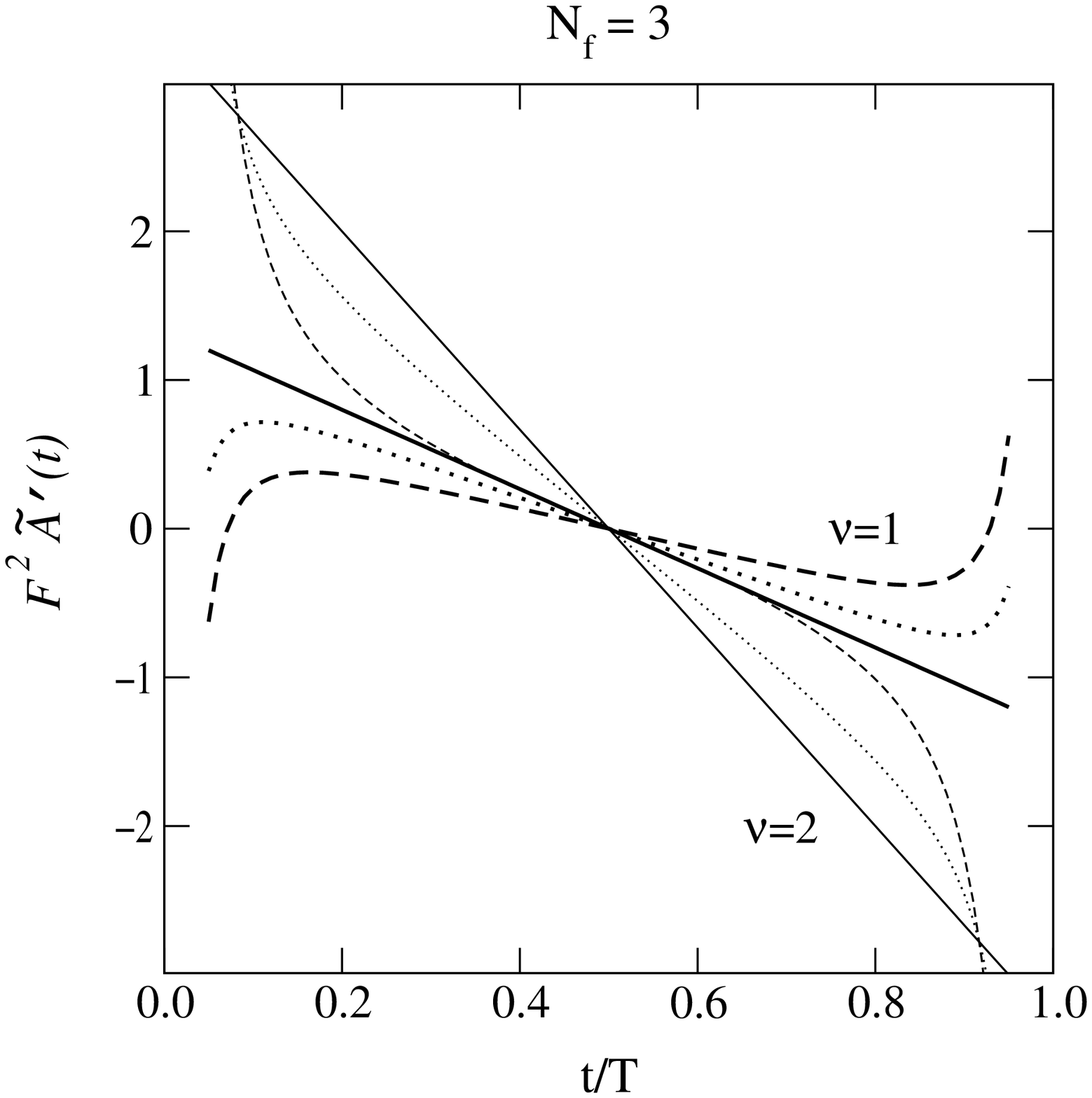,width=7.7cm}%

\caption[a]{The NLO (solid, volume-independent) and 
NNLO (dashed for $L=T=2$ fm, dotted for $L=T=3$ fm) predictions 
for $F^2 A'(t)$, $F^2 \tilde A'(t)$ 
at $|\nu|=1,2$ (thick, thin),  for $\Nf=2$ (top) and $\Nf=3$ (bottom). 
We have chosen $F = 93$~MeV.}
\la{fig:full}

\end{figure}

Fig.~\ref{fig:full} shows the NLO and NNLO results 
for $F^2 A'(t)$, $F^2 \tilde A'(t)$ 
as a function of time, for $\Nf=2$ and $\Nf=3$ and two volumes. 
Considering, say, the slope of the curves at around $t/T = 0.5$, 
the NNLO correction is $\sim 50$\% of the NLO term in the 
smaller volume shown, if $|\nu|$ is not too large, and then decreases
in larger volumes as $\sim 1/\sqrt{V}$.

%

\section{Quenched lattice determination of the low-energy couplings}
\la{sec:feasibility}

In this section, we move on from the full theory, which at present is not
easily accessible to lattice techniques, to consider
its quenched approximation. We derive the quenched chiral perturbation 
theory~\cite{BG,S} predictions for the pseudoscalar 
correlation functions of zero-mode eigenfunctions and compare them with 
numerical results obtained in quenched QCD with the 
overlap Dirac operator.

\subsection{Correlators in quenched chiral perturbation theory}
\la{cqxpt}

The predictions obtained with $\chi$PT, \eqs\nr{nonsinglet} and \nr{singlet},
diverge in the formal limit $\Nf \to 0$. This indicates that the 
results will be substantially modified in the quenched theory. 
Our working hypothesis is that correlators of the form of
\eq\nr{correlators} can nevertheless, at large distances
and in a certain kinematical range,
still be described by an effective chiral theory, 
called quenched chiral perturbation theory (Q$\chi$PT).  

The most important difference between
Q$\chi$PT and $\chi$PT is that the singlet field $\Phi_0
\sim \ln \det U$ cannot be integrated out in Q$\chi$PT~\cite{BG,S}.  
The corresponding chiral Lagrangian may then contain all possible
couplings of the singlet field and the theory 
loses much of its predictive power, 
unless an additional expansion in $1/N_c$ is carried out. 
In this case, the analysis
of the relevant operators follows very closely the
analysis of the generalized chiral theory,
including the $\eta'$ in full QCD~\cite{largenc}--\cite{l}, and is reviewed 
in~\ref{se:largeNc}. The presence of new couplings implies that
$\epsilon$-counting rules have to be defined for them. There are
several possibilities, as we also discuss in~\ref{se:largeNc}. 
We choose one that has not been considered previously, to our knowledge,
for reasons that will presently become clear. 

In the so-called supersymmetric formulation, the quenched chiral 
Lagrangian at the order we are working reads 
\ba
 \mathcal{L}_\rmi{Q$\chi$PT}  &=& 
 \frac{F^2}{4} \str 
 \Bigl[ \partial_\mu U \partial_\mu U^{-1} \Bigr] 
 - {m \Sigma \over 2} \str
 \! \Bigl[ U_{\theta} U + U^{-1} U^{-1}_{\theta}\Bigr] \nn 
 & - & i m K \Phi_0 \str
 \! \Bigl[ U_{\theta} U - U^{-1} U^{-1}_{\theta}\Bigr]
 + {m_0^2 \over 2 N_c} \Phi^2_0 + \frac{\alpha}{2N_{c}}
 (\partial_{\mu} \Phi_0)^2
 \;, \la{Lqchpt}
\ea
where $U \in \widehat{\rm Gl}(\Nv | \Nv)$~\cite{Z},
$\str$ denotes the supertrace, 
$\Phi_0 \equiv \frac{F}{{2}} 
\str [-i \ln(U)]$ and the vacuum angle $\theta$
appears as $U_\theta\equiv
\exp(i \theta {I_{\Nv}} /\Nv)$, where $I_{\Nv}$ is now the identity 
in the physical $\Nv\times \Nv$ ``fermion--fermion'' block and zero otherwise. 
The matrix $U_\theta$ commutes with all the flavour group
generators $T^I$, which are also assumed 
to be extended to become $2 \Nv \times 2 \Nv$ matrices, 
with only the physical block non-trivial.
Besides $F, \Sigma$, the quenched Lagrangian in~\eq\nr{Lqchpt} 
contains now three additional parameters: 
$K$, $m_0^2/N_c$ and $\alpha/N_c$. 
At the same order as \eq\nr{Lqchpt}, 
the operator corresponding to \eq\nr{SP} becomes
\ba
 {P}^I & = & 
 \;\, i \frac{\Sigma}{2} \str \Bigl[ T^I 
  \Bigl( U_\theta U - U^{-1} U_\theta^{-1} \Bigr)\Bigr]
  - \; K \Phi_0 \str \Bigl[ T^I 
  \Bigl( U_\theta U + U^{-1} U_\theta^{-1} \Bigr)\Bigr]
 \;. \la{cqPa}
\ea  

In the $\epsilon$-counting we have adopted, the mass parameter
related to the singlet field, $m_0^2/N_c$, will be treated as a small 
quantity of $\rmO( \epsilon^4)$, so that only the first order in it 
needs to be accounted for. The reason is that this 
guarantees that the non-zero mode Gaussian 
integrals over the graded group, performed
according to Zirnbauer's prescription~\cite{Z,ddhj}, are formally  
well defined. This counting also automatically implies that 
\be
  \frac{1}{(4 \pi F)^2} \ll \sqrt{V} \ll \frac{(4\pi)^2 N_c}{m_0^2}
  \;, \la{Qwindow}
\ee
which is the window where Q$\chi$PT should converge. 
Indeed,  quenched corrections increase in size with the volume 
in contrast with the unquenched case where they decrease: 
contributions of the form  $m_0^2 \sqrt{V}/N_c\sim \langle \nu^2 
\rangle/\sqrt{V} F^2$ become large if we do not satisfy \eq\nr{Qwindow}.
In the real world, obviously, $1/N_c$ is not tunable, and 
a phenomenological justification for the counting introduced is simply
that it seems to be able to describe our data,
as shown in the next sections.

Correlators of the form of~\eq\nr{correlators} again factorize
into two types of pieces, space-time integrals and zero-mode integrals. 
In the quenched 
theory the zero-mode integrals can only have terms 
$\propto \Nv$ (from the connected contraction) and $\propto \Nv^2$
(from the disconnected one). The connected contraction 
then directly determines
the result for the non-singlet correlator. The two parts can 
be determined as discussed in~\cite{currents}: the former by using 
the replica formulation~\cite{ds} 
and the U($\Nf$) integrals that already appeared in 
the full theory, the latter by carrying out the full computation 
of the zero-mode integrals 
for $\Nv = 1$, and subtracting the connected part. Therefore,
it is enough to consider $C^{00}_\nu$,
for a general $\Nv$, and 
deduce $C^{aa}_\nu$ from the part $\propto \Nv$ in $C^{00}_\nu$.

Generalizing \eq\nr{Cx}, the overall form of the answer now is
\ba
 C^{00}_\nu(x) \!\!\! & = & \!\!\!  
 C_0 + 
 \alpha_0 G(x) + \alpha_0' E(x)
 + \beta_0\Bigl[ G(x) \Bigr]^2 
 + \beta_0' G(x) E(x) 
 + \beta_0''\Bigl[ E(x) \Bigr]^2  
 \la{qCx}
 \\ \!\!\! & + & \!\!\!  
 \int \! {\rm d}^4 y \,
 \Bigl[ 
 \gamma_0 G(x-y) G(y)
 + \gamma_0' G(x-y) E(y)
 + \gamma_0'' E(x-y) E(y)
 \Bigr]
 + \epsilon_0 \delta^{(4)}(x) 
 \;, \nonumber
\ea
where, instead of $E(x)=G(x)/\Nf$ as in the unquenched theory, 
we now have
\ba
 E(x) 
 & \equiv & \frac{\alpha}{2 N_c} G(x) + \frac{m_0^2}{2 N_c} F(x)
 \;.
 \la{qprop}
\ea
Here $G(x)$ is defined in~\eq\nr{Gx}, and
\be
 F(x) = \frac{1}{V} 
 \sum_{n \in \mbox{\raisebox{-0.2ex}
 {$\scriptstyle\mathsf{Z\hspace*{-1.3mm}Z}^4$}}} 
 \Bigl(1 - \delta^{(4)}_{n,0} \Bigr) \frac{e^{i p \cdot x}}{p^4} 
 \;, \quad
 p = 2\pi\Bigl(\frac{n_0}{T}, \frac{\vec{n}}{L} \Bigr)
 \;.
\ee

The additional time-dependent functions appearing in the quenched
case, owing to the function $F(x)$ in~\eq\nr{qprop}, are listed in
Appendix~\ref{app:qints}. The quenched zero-mode integrals are
discussed in Appendix~\ref{app:qzeromode}, and the expressions
for the coefficients in \eq\nr{qCx}, in terms of the zero-mode
integrals, in Appendix~\ref{app:qcoeffs}.

Collecting everything together, we obtain for the objects 
in~\eqs\nr{defA}, \nr{deftA}, 
\ba
 F^2 A'(t)
 & = & 
 2 |\nu|
 \biggl[ 
 |\nu|  h_1' (\tau) 
 + \biggl( {\alpha \over 2 N_c} 
 - \frac{2 K F}{ \Sigma} 
 - {\beta_1 \over F^2 \sqrt{V}} 
  \biggr)  h_1' (\tau) 
 \nn & &   
  + \frac{T^2}{F^2 V} \biggl( 2 \nu^2 + \fr73  
  - 2 \langle\nu^2\rangle \biggr) h_2'(\tau)   
  + \frac{T^2}{2 F^2 V} g_1'(\tau) 
 \biggr] 
 \;, \la{zeromode_conn_NLO_improved} \\
 F^2 \tilde A'(t) 
 & = & - 2 |\nu|
 \biggl[ 
  h_1' (\tau) 
  +|\nu| \biggl( {\alpha \over 2 N_c} 
 - \frac{2 K F}{ \Sigma} 
 - {\beta_1 \over F^2 \sqrt{V}}
 \biggr)  h_1' (\tau) 
 \nn & & 
 +  
 \frac{T^2}{F^2 V} \biggl( \fr{13}3 |\nu| - 2 |\nu| \langle\nu^2\rangle
  \biggr)  h_2'(\tau)   
  +|\nu| 
 \frac{T^2}{2 F^2 V} g_1'(\tau) 
 \biggr] 
 \;. \hspace{0.5cm}
 \la{zeromode_disconn_NLO_improved} 
\ea
We have used here the Witten--Veneziano relation 
$m_0^2 F^2 = 4 N_c \langle \nu^2\rangle/V$, which is exact at this 
order, where $\langle \nu^2 \rangle/V$ is the topological susceptibility.  
It may be noted that for $|\nu|=1$, 
$A'(t) + \tilde A'(t) = 0$, as 
should be the case. We observe that 
there are three independent low-energy parameters entering the 
expressions: $\langle \nu^2\rangle$, 
the combination $\alpha/2 N_c- 2 K F/\Sigma$, and $F$; we thus set, 
without loss of generality, $K=0$. Obviously 
a simultaneous determination of three parameters from 
the zero-mode eigenfunctions 
will be more difficult than in the unquenched case, 
where only $F$ appears at this order. 

\begin{figure}[t]
\begin{center}

\epsfig{file=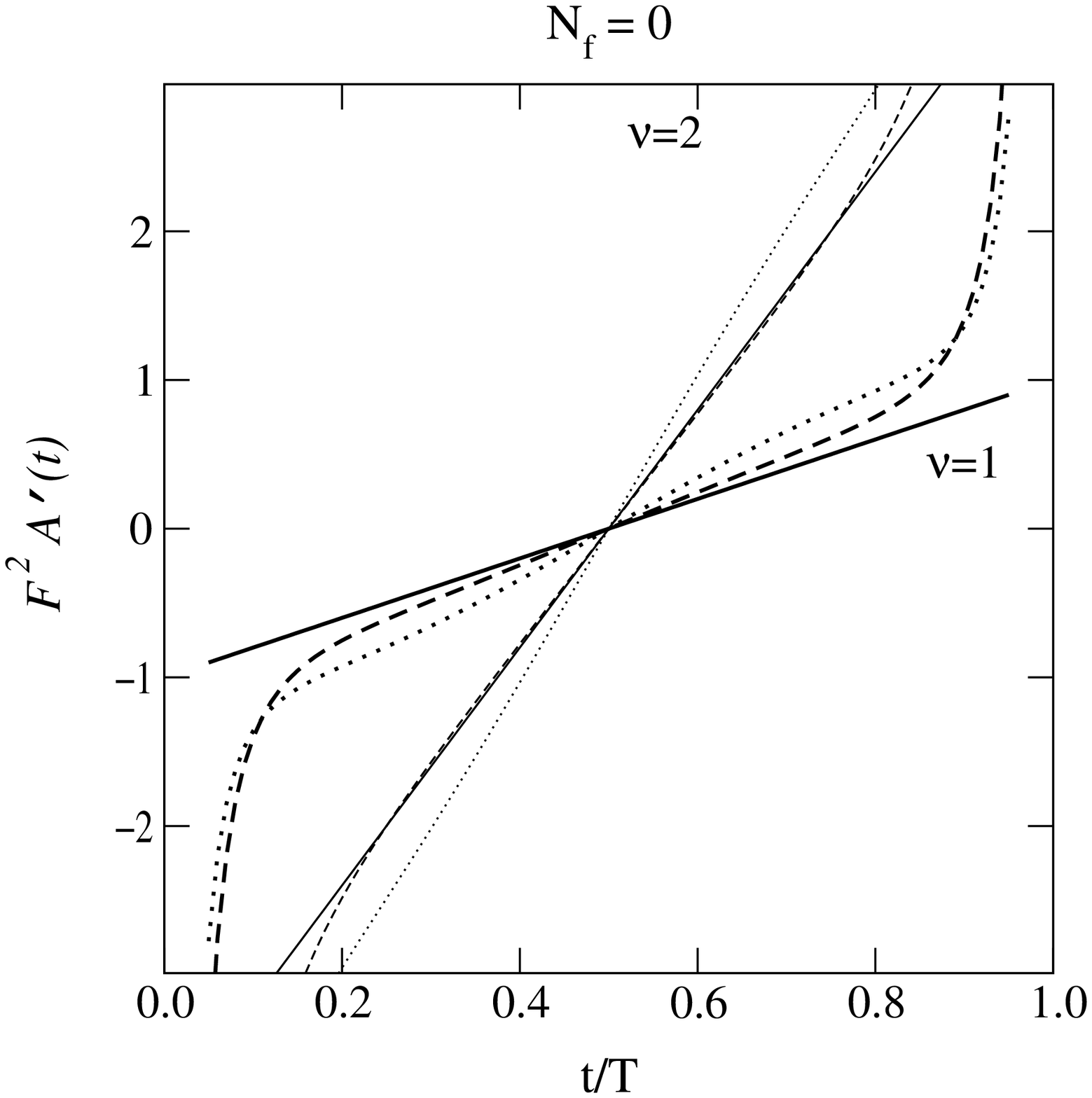,width=7.7cm}%
\epsfig{file=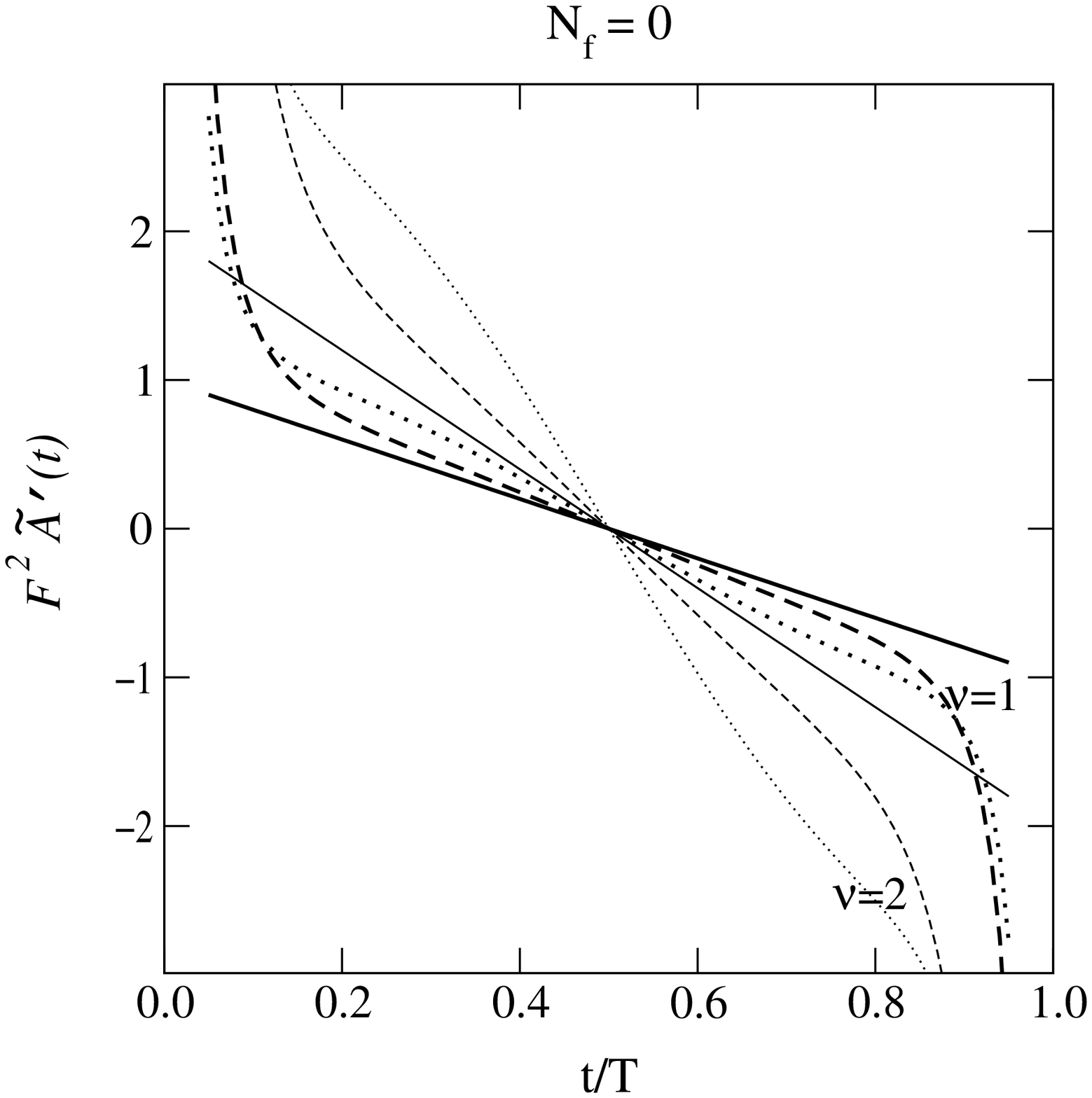,width=7.7cm}%

\caption[a]{The NLO (solid, volume-independent) and
NNLO (dashed for $L=T=1.6$~fm, dotted for $L=T=2.0$~fm)
predictions from \eqs\nr{zeromode_conn_NLO_improved}, 
\nr{zeromode_disconn_NLO_improved}. The Q$\chi$PT 
parameters are $\alpha/N_c=0$, $F = 115$~MeV, 
$\langle \nu^2 \rangle/V = (200~\mbox{MeV})^4$, 
and $|\nu|=1,2$ (thick, thin).}
\la{fig:quenched}

\end{center}
\end{figure}

In Fig.~\ref{fig:quenched} we show the NLO and NNLO predictions
for $F^2 A'(t)$ and $F^2 \tilde A'(t)$, 
\eqs(\ref{zeromode_conn_NLO_improved}) and 
(\ref{zeromode_disconn_NLO_improved}), for $L= 1.6$ and 2.0 fm.
Considering, say, the slope of the curves at around $t/T = 0.5$, 
the NNLO correction grows to $\sim 50$\% of the NLO term 
in the larger volume.

\subsection{Simulation details}

We have performed a lattice simulation in the quenched approximation, using 
the overlap Dirac operator for the fermions \cite{hn}. The topological 
index and the zero-mode eigenfunctions are computed 
as proposed in~\cite{methods} on thermalized 
configurations for two physical volumes and various lattice spacings.
Only sectors with topology $|\nu|=1,2$ are considered.
Table~\ref{tab:data} summarizes the simulation parameters; the 
same configurations have previously 
been analysed in a different context~\cite{glww}. 

\begin{table}
\begin{center}
\begin{tabular}{lllllll}
\hline
Lattice & ~~~$\beta$ & $L/a$ & ~$r_0/a$ & $L$[fm] & 
$N_\rmi{meas}(|\nu|=1)$ & $N_\rmi{meas}(|\nu|=2)$ \\
\hline
~~~B$_{0}$ & 5.8458 & ~12 & 4.026 & ~1.49 & ~~~~~~880 & ~~~~~~696 \\
~~~B$_{1}$ & 6.0    & ~16 & 5.368 & ~1.49 & ~~~~~~307 & ~~~~~~226 \\
~~~B$_2$   & 6.1366 & ~20 & 6.710 & ~1.49 & ~~~~~~326 & ~~~~~~213 \\
~~~C$_0$   & 5.8784 & ~16 & 4.294 & ~1.86 & ~~~~~~229 & ~~~~~~186 \\
~~~C$_1$   & 6.0    & ~20 & 5.368 & ~1.86 & ~~~~~~83  & ~~~~~~78 \\
\hline
\end{tabular}
\caption{The simulation parameters (cf.\ ref.~\cite{glww}). Here 
$a$ is the lattice spacing, 
$r_0$ is the Sommer scale~\cite{ss}, $r_0 = 0.5$~fm, 
and $N_\rmi{meas}$ is the number of configurations.
All lattices are symmetric, $T=L$.}
\la{tab:data}
\end{center}
\end{table}

\begin{figure}[t]
\begin{center}

\epsfig{file=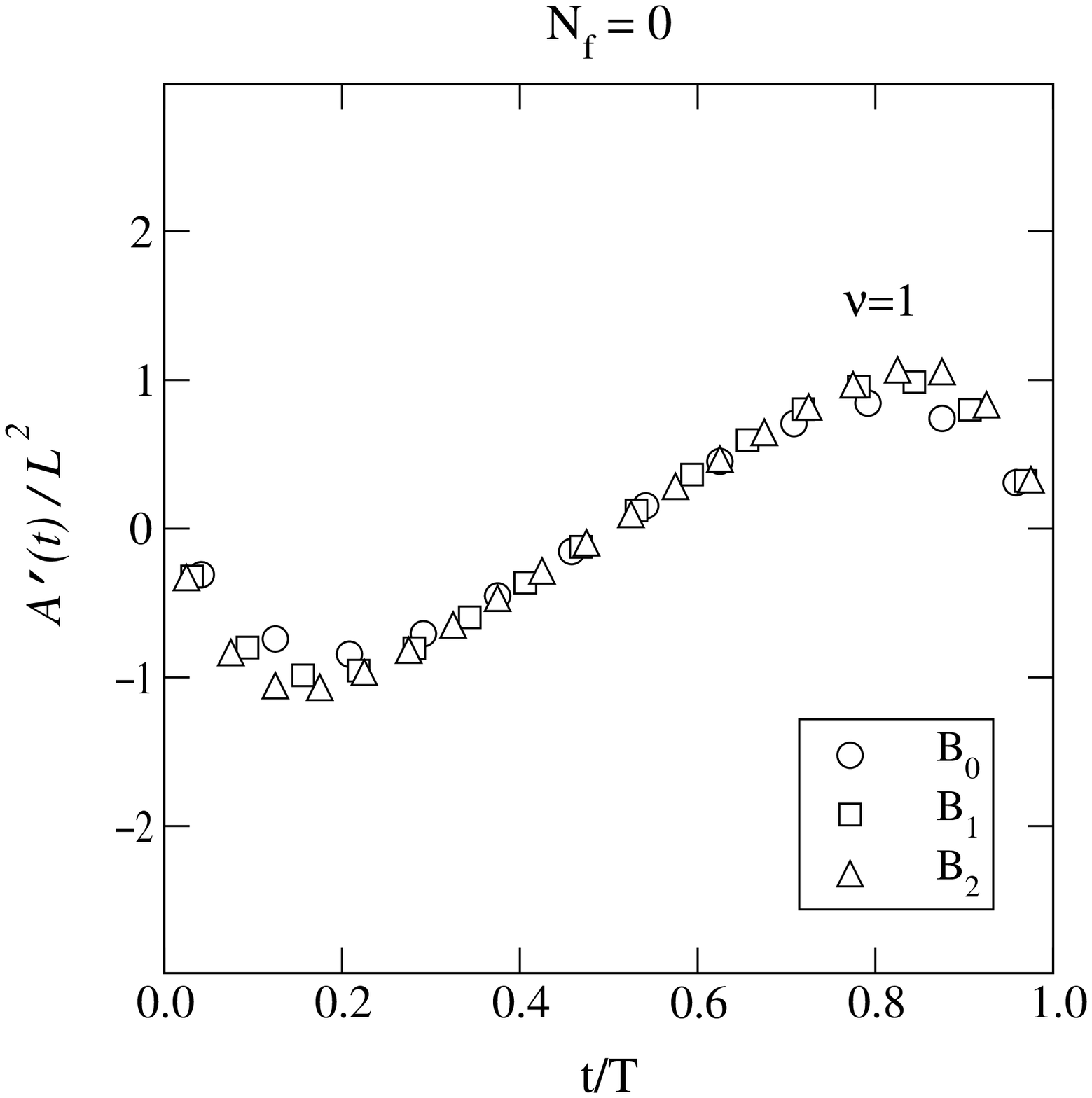,width=7.7cm}%
\epsfig{file=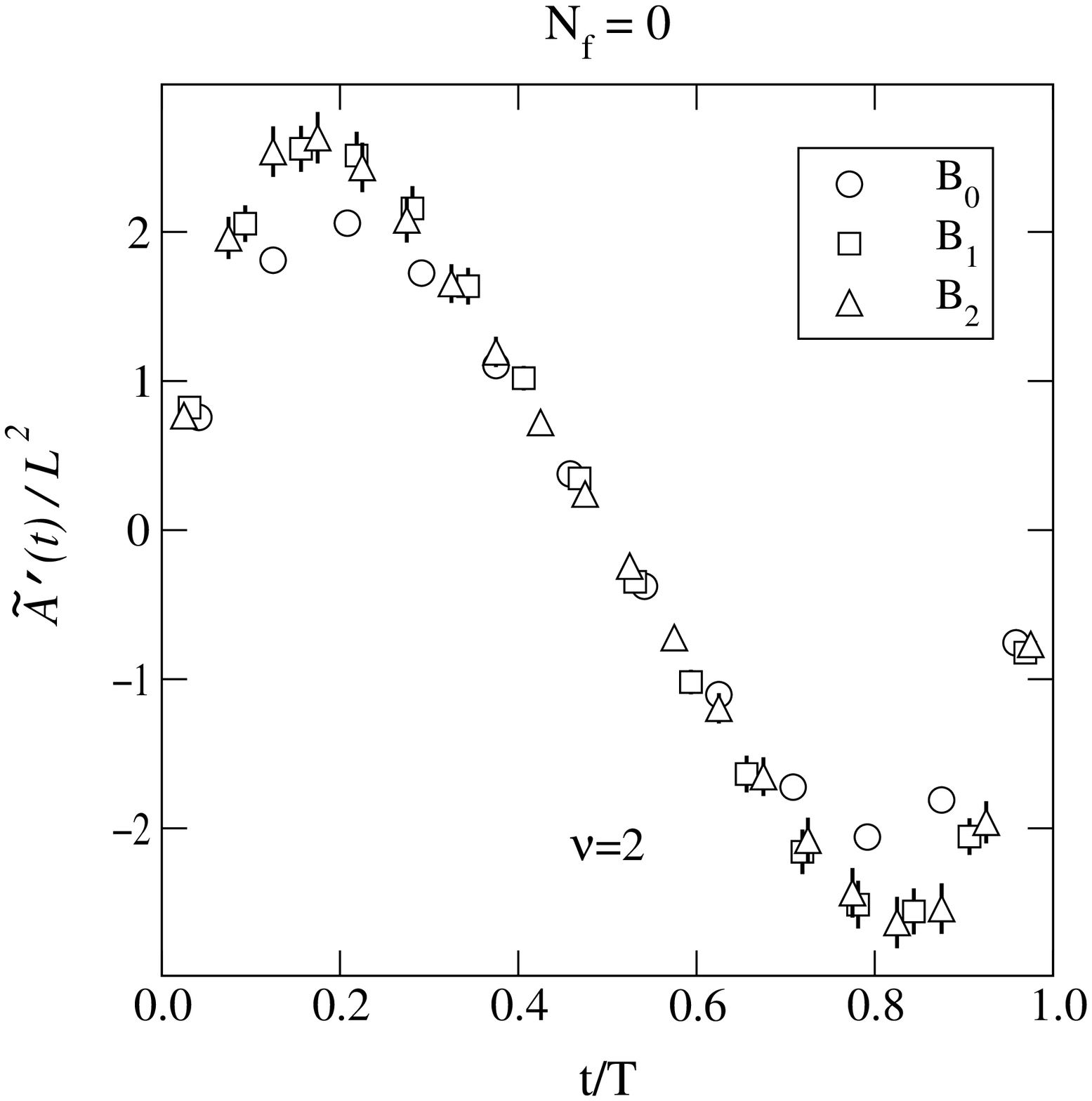,width=7.7cm}%

\caption[a]{The numerical data corresponding to
$\mathcal{A}'(t)/L^2$, $\tilde \mathcal{A}'(t)/L^2$, from the B lattices.
Where not visible the statistical errors are smaller than the symbols.
The left plot is for $|\nu|=1$, the right one for $|\nu|=2$. 
Comparing the slope at $t/T = 0.5$ with the $\chi$PT predictions shown
in~\fig\ref{fig:quenched}, allows us to estimate $(FL)^2$.} 
\la{fig:data}

\end{center}
\end{figure}

{}From the zero-mode eigenfunctions, 
we compute the volume average of the correlators
in \eqs(\ref{eq:v}) and (\ref{eq:vbar}). 
There is a good signal in all cases, 
as illustrated in~\fig\ref{fig:data}. 
Q$\chi$PT predicts that at non-zero times
these correlators should behave like polynomials
in time. We thus consider a Taylor expansion around 
the mid-point, $\tau = 1/2$. Denoting $z\equiv \tau - 1/2$, 
we define the coefficients $D_\nu$ and $C_\nu$ as 
\ba
 \frac{1}{L^2} \mathcal{A}'(t) & \equiv & 
 D_\nu \, z + C_\nu z^3 + \rmO(z^5)
 \;, \la{Dnu} \\
 \frac{1}{L^2} \tilde \mathcal{A}'(t) & \equiv & 
 \tilde D_\nu \, z + \tilde C_\nu z^3+ \rmO(z^5)
 \;. \la{Dnup}
\ea
With a simple linear fit we can then 
extract the parameters $D_\nu$ and $C_\nu$ 
on jackknifed configurations. Table~\ref{tab:dcs} shows the results of 
these fits in the time interval $(t_{\rm min}, T/2)$. The data 
are modelled very well by the fits, and also the dependence 
on the choice of $t_{\rm min}$ is insignificant. 

It is clear from Table~\ref{tab:dcs} that only the $D_\nu$
coefficients can be extracted from the data in a reliable way. 
The errors on the $C_\nu$ coefficients
are large and their central values vary quite significantly with the 
lattice spacing. This is to be expected since the $C_\nu$ coefficients are
more relevant at short distances and so will also  
be more sensitive to cutoff effects. For this reason, we restrict
ourselves to the $D_\nu$ coefficients in the following.

\begin{table}
\begin{center}
\begin{tabular}{llllllrrl}
\hline
Lattice & $t_{\rm min}$ & $z_\rmi{min}$  & 
  ~~$D_1$  & ~~~$D_2$ & ~~~~~$\tilde D_2$ &
    $C_1$~ & $C_2$~~ & ~$\tilde C_2$~ \\
\hline
~~~B$_0$ & ~$3a$ & $0.208$
                 & $3.7(1)$
                 & $10.9(2)$
                 & $-\phantom{1}9.1(4)$
                 & $-\phantom{1}7(1)$
                 & $-24(1)$
                 & $19(3)$    \\
~~~B$_1$ & ~$4a$ & $0.219$
                 & $3.9(2)$
                 & $10.8(4)$
                 & $-11.1(9)$
                 & $-\phantom{1}5(2)$
                 & $-11(3)$
                 & $26(8)$    \\
~~~B$_2$ & ~$5a$ & $0.225$
                 & $3.8(3)$
                 & $10.7(4)$
                 & $-\phantom{1}9.7(9)$
                 & $-\phantom{1}3(3)$
                 & $-13(4)$
                 & $\phantom{1}8(8)$   \\
~~~C$_0$ & ~$4a$ & $0.219$
                 & $3.4(3)$
                 & $\phantom{1}8.9(5)$
                 & $-\phantom{1}9.2(9)$
                 & $\phantom{-1}\,2(3)$
                 & $-\phantom{1}4(5)$
                 & $16(9)$   \\
~~~C$_1$ & ~$5a$ & $0.225$
                 & $3.9(5)$
                 & $\phantom{1}9.4(7)$
                 & $-\phantom{1}9.1(16)$
                 & $-11(5)$
                 & $\phantom{-1}7(7)$
                 & $\phantom{1}2(19)$   \\
\hline
\end{tabular}
\caption{The $D_\nu$ and $C_\nu$ coefficients from the fit. The numbers
 in parentheses indicate the error of the last digit.}
\la{tab:dcs}
\end{center}
\end{table}

\subsection{Analysis of the data}

A Taylor expansion of the functions 
in \eqs(\ref{zeromode_conn_NLO_improved}) 
and (\ref{zeromode_disconn_NLO_improved}) and
a matching with \eqs\nr{Dnu} and \nr{Dnup} gives
\ba 
 D_\nu & = & 
 \frac{2|\nu|}{(FL)^2}
 \biggl\{
 |\nu|  + \frac{\alpha}{2 N_c} - \frac{2 K F}{\Sigma}
 - \frac{\beta_1}{F^2\sqrt{V}} 
 \nn & & 
 \hphantom{ \frac{2|\nu|}{(FL)^2} \biggl\{ }
 + \biggl[ 
 \biggl( 
 \fr73 + 2 \nu^2 - 2 \langle \nu^2 \rangle
 \biggr) \zeta_2 + \fr12 \gamma_1
 \biggr] \frac{T^2}{F^2 V} 
 \biggr\} 
 \;, \la{qDnu} \\
 \tilde D_\nu & = & 
 \frac{2|\nu|}{(FL)^2}
 \biggl\{
 -1 - |\nu| \biggl( \frac{\alpha}{2 N_c} - \frac{2 K F}{\Sigma} 
 - \frac{\beta_1}{F^2 \sqrt{V}} 
 \biggr) 
 \nn & & 
 \hphantom{ \frac{2|\nu|}{(FL)^2} \biggl\{ }
 - |\nu| \biggl[ 
 \biggl( \frac{13}{3} - 2 \langle \nu^2 \rangle \biggr)\zeta_2 + 
 \fr12 \gamma_1 
 \biggr] \frac{T^2}{F^2 V}
 \biggr\} \la{qDnup}
 \;,
\ea
where we have written
$h_2'(\tau) = \zeta_2 z + \rmO(z^3)$, 
$g_1'(\tau) = \gamma_1 z + \rmO(z^3)$, and 
\be
 \zeta_2 =  -\frac{1}{24}  \;,\;\; 
 \gamma_1 =  -\frac{1}{12} + \fr12
 \sum_{{\vec n} \neq 0} \frac{1}{{\sinh}^2\left(|{\vec p}|/2\right)}
 \approx   -0.0571276522\;, ~~~\mbox{for $T=L$}
 \;,
\ee
with $|{\vec p}| = 2 \pi  T [ \sum_{i=1}^3  n_i^2 ]^{1/2} /{L}$.

The quantity $\langle \nu^2\rangle$ 
in~\eqs\nr{qDnu} and \nr{qDnup}
has recently been computed 
with high accuracy~\cite{glww,ddp}. In ref.~\cite{glww} 
the results obtained at several lattice spacings were consistent
with a well-defined continuum limit~\cite{grtv}, giving
$\hat\chi \equiv r_0^4 \langle \nu^2 \rangle/V = 0.059(5)$, 
where $r_0 = 0.5$~fm~\cite{ss}.
In order to determine the other parameters $F,\alpha/N_c$, 
we need to fit for them
simultaneously, but also take into account the
error in the determination of $\hat\chi$.  By comparing
the results for the $D_\nu$ coefficients on the different
lattices, cutoff effects are seen to be negligible within the 
statistical uncertainty. For this reason we do not attempt a continuum
extrapolation here and simply consider the data at different lattice
spacings as statistically independent. Since, on the other hand, the 
value of $\hat\chi$ cited above is the result of
a continuum extrapolation,
we will assign new error bars to it, large enough to also 
incorporate the finite lattice spacing values from~\cite{glww}:
$\hat\chi_0 \equiv 0.059(10)$. We then perform a 
$\chi^2$ minimization in the three-parameter space $(F,\alpha/N_c,\hat\chi)$,
with $\hat\chi$ added to the $\chi^2$ function 
as $\delta \chi^2 = [(\hat\chi - \hat\chi_0)/\delta \hat\chi_0]^2$.

We have performed three fits: the B lattices, the C lattices and their
combination, taking into account the ratio of physical volumes,
$L^\rmi{({B})}/L^\rmi{({C})}= 16/20$. 
The values of $\chi^2_{\rm min}$ 
and the projections of the 68\% confidence level contours onto the 
different parameter axes can be found in 
Table~\ref{tablefits}. The quality of
the fits is good, with $\chi^2_{\rm min}/\mbox{dof} \lsim 1.0$ 
in all cases, and the B and C lattices give rather compatible results. 

\begin{table}
\begin{center}
\begin{tabular}{llllll}
\hline
Lattice & dof & $\chi^2_{\rm min}$ & ~~~~$F L^\rmi{({B})}$ & 
 ~~~~~$\alpha/N_c$ & 
 ~~~~~~$\hat \chi$   \\
\hline
~~~B & ~~6 & 5.8 & (0.84,0.92) & ~~\,(0.3,1.2) &  (0.04,0.08) \\
~~~C & ~~3 & 2.3 & (0.76,1.01) & ($-$0.6,2.6) &  (0.04,0.08) \\
~B+C & ~12 & 8.8 & (0.83,0.90) & ~~\,(0.3,0.9) &  (0.05,0.08) \\
\hline
\end{tabular}
\caption{Results from the global fits.
The intervals are the projections of the
68\% confidence level contours.
\label{tablefits}}
\end{center}
\end{table}

It is interesting to contrast this situation with 
what it would be in full QCD, where 
the $D_\nu$ coefficients only depend on the decay constant~$F$:  
\ba 
 D_\nu & = & 
 + \frac{2|\nu|}{\Nf (FL)^2}
 \biggl\{
 (1  + |\nu| \Nf) \biggl(1 - \Nf \frac{\beta_1}{F^2\sqrt{V}}\biggr)
 +\frac{T^2}{F^2 V}  \biggl[\gamma_1 \biggl(\frac{2 + \Nf^2}{2 \Nf}
 + {4-\Nf^2 \over 2} |\nu| \biggr) 
 \nn &+&  
 \zeta_2 \biggl( (6-\Nf^2) |\nu| + {4\over \Nf} 
 + \Nf (2\nu^2-1) \biggr) \biggr] \biggr\}
 \;, \la{Dnufull} \\
 \tilde D_\nu & = & 
 -\frac{2|\nu|}{\Nf (FL)^2}
 \biggl\{
 (\Nf +|\nu|) \biggl(1 - \Nf \frac{\beta_1}{F^2 \sqrt{V}} \biggr)
 + \frac{T^2}{F^2 V}  \biggl[\gamma_1 \biggl({4-\Nf^2\over 2}+ 
  \frac{2 + \Nf^2}{2 \Nf} |\nu| \biggr) 
 \nn &+& 
 \zeta_2 
 \biggl(
 4 + 2 \nu^2 - \Nf^2 + \frac{4}{\Nf}|\nu| + \Nf |\nu| 
 \biggr)
 \biggr] \biggr\}
 \;. \la{tDnufull}
\ea
These expressions show reasonable convergence (in the sense 
that the NNLO correction is less than 50\% of the NLO term) 
only at $L \gsim 2.0$~fm for $F=93$~MeV, and in order to push the size 
of the correction below 30\%, one would need to go to $L \gsim 2.5$~fm.
In this case we might expect a systematic uncertainty in the determination
of $F$ of about 5\%, and statistical uncertainties 
would reach the same level if $D_\nu$ could be determined 
using $\sim 100$ configurations with $|\nu|=1$.

\begin{figure}[t]
\begin{center}

\epsfig{file=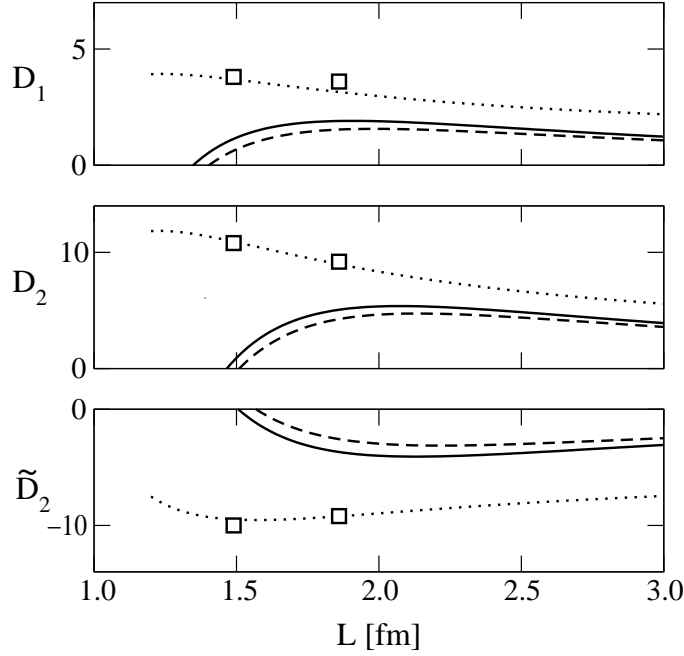,width=9cm}

\caption[a]{The coefficients $D_1$, $D_2$ and $\tilde D_2$ 
as a function of the box size
for $\Nf=2$ (solid), $\Nf=3$ (dashed) and quenched (dotted). 
The quenched parameters are chosen at their best fit values
according to Table~\ref{tablefits}, while 
$F = 93$~MeV for $\Nf = 2,3$. The symbols denote the 
averages of the data points from the B ($L^\rmi{({B})}=1.49$~fm) 
and C ($L^\rmi{({C})}=1.86$~fm) lattices.}

\la{ds}

\end{center}
\end{figure} 

The full theory formulae, \eqs\nr{Dnufull}, \nr{tDnufull}, 
possess the feature that the NNLO corrections 
come with negative relative signs, such that the expressions are almost
independent of $F L$ at around $F L \sim 1$, and their 
absolute values have an upper bound at this order. For illustration, 
we show the full predictions for $F = 93$~MeV  in Fig.~\ref{ds}, 
as a function of the box size (for a symmetric geometry, $L=T$). 
Also shown are the quenched data as well as the quenched predictions. 
Because of the mentioned near-cancellation, the full predictions 
at this order could not be moved significantly
closer to the quenched data points by tuning $F$. 
In any case, as already mentioned, they show reasonable
convergence and can thus be considered self-consistent predictions 
only for $L \gsim 2.0$~fm.

%
\section{Conclusions}
\la{sec:conclu}

Approaching the chiral limit has remained a long-standing challenge 
for lattice QCD for many reasons, among them that finite-volume effects
become large for very light pseudo-Goldstone bosons, and that the Dirac
operator develops very small eigenvalues. 
It has been the purpose of this paper to elaborate on the fact 
that at least these particular problems can be overcome: 
for instance, the Dirac operator eigenfunctions associated with 
the exact zero modes encountered
in gauge field configurations with a non-trivial 
topology at finite volume, can be used to extract physical information 
concerning the chiral limit of the infinite-volume theory. 

More precisely, we have shown
that certain classical scattering amplitudes of 
the zero-mode eigenfunctions measured at finite volumes, 
\eqs\nr{eq:v}, \nr{eq:vbar}, allow the extraction of the 
infinite-volume pion decay constant, 
via the relations in \eq\nr{relation}.
We have worked out these relations to NNLO, 
\eqs\nr{nonsinglet}, \nr{singlet}, finding that 
the convergence of chiral perturbation theory  
seems reasonable for these observables, 
provided the volume is above $\sim$ (2.0 fm)$^4$.

Finally, to estimate the practical feasibility of using such 
relations, we have carried out lattice Monte Carlo simulations 
in the quenched approximation, using overlap fermions. We find 
a good signal for the observables in~\eqs\nr{eq:v}, \nr{eq:vbar}, 
shown in \fig\ref{fig:data}. 
Matching with chiral perturbation theory predictions relevant
to the quenched approximation (which show reasonable apparent
convergence only in volumes between $\sim$ (1.0 fm)$^4$ and 
$\sim$ (2.0 fm)$^4$, 
in marked contrast with the unquenched case), 
we find that the pion decay constant, 
to the extent that it is a well-defined quantity in this case, 
can be extracted with about 5\% statistical
accuracy, utilizing a few hundred 
configurations with non-trivial topology. 
The number we obtain in volumes $\sim$ (1.5 fm)$^4$
is in the ballpark of 115~MeV.

Our result for the pion decay constant in the chiral limit is 
larger than what one would expect in Nature: conventional $\chi$PT in
infinite volume~\cite{gl2} yields $F\approx 87$~MeV, if the standard
phenomenological values for the $\rmO(p^4)$ $L_i$ coefficients of 
Gasser and Leutwyler are
inserted~\cite{BijEckGas94}. The fact that our quenched calculation
seems to overestimate $F$ is consistent with other recent quenched
results for the physical $F_\pi$ in the continuum 
limit, however~\cite{ALPHA_ms}--\cite{CP_PACS_quen}. 
For instance, the results of~\cite{ALPHA_ms,ALPHA_Leff} imply 
that the quenched $F_\pi$ is 10\% larger than the experimental 
value, if the scale is set by $r_0$~\cite{ss}. On the other hand, 
these standard approaches (unlike ours) have to rely on quenched 
chiral extrapolations in the light quark masses, which introduce 
significant systematic uncertainties of their own~\cite{panel}. 

On the side of our approach, it is conceivable that 
a smaller value for $F$ could be obtained by going to larger volumes.
As we have discussed, however, the peculiarities of the quenched 
approximation imply that the volume cannot be increased too much, 
since the convergence of quenched chiral perturbation theory soon  
deteriorates. Therefore, a systematically improvable determination of $F$ 
by using our method (or any other) lies beyond the quenched approximation.

%
\section*{Acknowledgements}

The present paper is part of an ongoing project whose final goal
is to extract low-energy parameters of QCD from numerical simulations 
with GW fermions. The basic ideas of our approach were developed 
in collaboration with M. L\"uscher; we would like to thank him for
his input and for many illuminating discussions. 
We are also indebted to P.H.\ Damgaard, K.~Jansen and L.\ Lellouch for 
interesting discussions.
The simulations were performed on PC clusters 
at the University of Bern, at DESY-Hamburg,
at the Max-Planck-Institut f\"ur Physik in Munich, 
at the Leibniz-Rechenzentrum der Bayerischen Akademie der Wissenschaften, 
and at the University of Valencia.
We wish to thank all these institutions for support, 
and the staff of their computer centres for technical help.
L.~G.~was supported in part by the EU under contract 
HPRN-CT-2000-00145 (Hadrons / Lattice QCD), and 
P.~H.\ by the CICYT (Project No.\ FPA2002-00612) and 
by the Generalitat Valenciana (Project No.\ CTIDIA/2002/5).

\newpage


\appendix
\renewcommand{\thesection}{Appendix~\Alph{section}}
\renewcommand{\thesubsection}{\Alph{section}.\arabic{subsection}}
\renewcommand{\theequation}{\Alph{section}.\arabic{equation}}


%
\section{Large-$N_c$ counting in the $\epsilon$-regime}
\la{se:largeNc}

Large-$N_c$ counting in the context of chiral perturbation theory
has been analysed in detail in ref.~\cite{l}. 
The same general discussion goes through  
in the full and in the quenched theories, with the replacements
in the latter that $U^\dagger \to U^{-1}, e^{i\theta/\Nf}\to U_\theta,
\tr \to \str$. For simplicity, 
we will mostly use the notation of the unquenched theory here, 
indicating then the important 
point at which differences arise between the two cases.

In general, the chiral theory including the singlet is, at leading
order in the momentum expansion and to all orders in $1/N_c$, of the form
\ba
 \mathcal{L}_\rmi{$\chi$PT} \!\! & = & \!\! 
 V_0(\Phi_0) + V_1(\Phi_0) ~ \tr 
 \Bigl[ \partial_\mu U \partial_\mu U^{\dagger} \Bigr] 
 \nn \!\! & - & \!\!  
 \Bigl\{ 
  V_2  (\Phi_0) \tr \! \Bigl[ e^{i\theta/\Nf} U M \Bigr]
 +V_2^*(\Phi_0) \tr \! \Bigl[M^{\dagger}U^{\dagger} e^{-i\theta/\Nf} \Bigr]
 \Bigr\} 
 + V_3(\Phi_0) ~(\partial_{\mu} \Phi_0)^2 + ...\;,
 \la{Lrep} \hspace*{0.9cm}
\ea
where $\Phi_0 \equiv -i \frac{F}{2} \tr \ln U$ and 
$M=\mathop{\mbox{diag}}(m,m,...)$.

The Lagrangian in~\eq\nr{Lrep} contains an infinite number of
parameters, since the potentials $V_i(\Phi_0)$ are arbitrary
functions, with the only constraint from parity 
that $V_i(-\Phi_0) = V_i(\Phi_0)$, 
for $i=0,1,3$ and $V_2(-\Phi_0) = V_2^*(\Phi_0)$. It can be shown,
however, that they involve a specific power series in $1/N_c$
(\cite{largenc}--\cite{l}, and references therein). 
Noting that the field $\phi_0$ of~\cite{l} is 
$\phi_0 = 2 \Phi_0/(\bar F \sqrt{N_c})$ in our notation below, 
the structures arising are
\ba
 V_0(\Phi_0) & \equiv & 
 \fr12 {m_0^2 \over N_c} \Phi_0^2 
 + \rmO\left({\Phi_0^4 \over N^4_c}\right)
 \;, \\
 V_1(\Phi_0) & \equiv & 
 {N_c {\bar F}^2 \over 4} 
 + \rmO\left({\Phi_0^2 \over N^2_c}\right)
 \;, \\
 V_2(\Phi_0) & \equiv & 
 {N_c {\bar \Sigma} \over 2} + 
 i {\bar K \over \sqrt{N_c}} \Phi_0 + 
 \rmO\left({\Phi_0^2 \over N^2_c}\right)
 \;, \la{defKbar} \\
 V_3(\Phi_0) & \equiv & 
 {\alpha \over 2 N_c}  
 + \rmO\left({\Phi_0^2 \over N^4_c}\right)
 \;, 
\ea
where all parameters introduced ($m_0^2,\bar F,\bar \Sigma, \bar K,
\alpha$) are assumed not to scale with $N_c$. Inserting the specific
terms shown here into~\eq\nr{Lrep}, one obtains the theory up to
$\rmO(1/N_c)$. In the following, we denote
\be
 F^2 \equiv {\bar F}^2 N_c, \quad
 \Sigma \equiv {\bar \Sigma} N_c, \quad
 K \equiv \frac{\bar K}{\sqrt{N_c}}
 \;. \la{ss}
\ee

In order to define a formally consistent framework, it is convenient 
to now combine the momentum and $1/N_c$ expansions. Following~\cite{l}, 
we may choose
\be
 p^2 \sim \frac{1}{N_c} \sim \epsilon^2 \;.
 \la{pcount}
\ee
The defining property of the $\epsilon$-regime is that the pions 
are off-shell since the momenta are fixed by the size of the box, 
$p\sim 1/L \sim \rmO(\epsilon)$, while the quark
mass is small, such that
\be
 \mu \equiv m \Sigma V \lsim 1 \;.
\ee
Given~\eqs\nr{ss}, \nr{pcount}, we are thus led to the rule 
\be
 m \sim \epsilon^6 \;.
\ee
As usual, the field configurations are factorized into zero-momentum 
modes $U_0, \bar \Phi_0$ and non-zero modes $\xi, \tilde \Phi_0$, 
\be
 U(x) = U_\xi(x)  U_0 \;, \quad
 U_\xi(x) = \exp\left[\frac{2i\xi(x)}{F}\right] \;, \quad
 \Phi_0(x) = \tilde \Phi_0(x) + \bar \Phi_0 \;, 
\ee
where $\int\! {\rm d}^4 x\, \xi(x) = 0$ 
and $\tilde \Phi_0 = \tr \xi$. The counting rules for the non-zero modes,
which are treated perturbatively, are
\be
 \xi \sim p \sim \epsilon \;, \quad
 \tilde \Phi_0 \sim p \sim \epsilon \;. 
\ee
For the zero mode $U_0$ we have
$U_0 \sim 1$, while the counting of $\bar \Phi_0$ is
to be determined presently. 

Indeed, let us consider the terms involving explicitly
the flavour singlet zero mode $\bar \Phi_0$. 
We are interested in carrying out the computation up to and including
$\rmO(\epsilon^4)$, and the terms potentially of this order,
after integration over space-time, are
\ba
  \int \! {\rm d}^4 x\, \mathcal{L}_\rmi{$\chi$PT} & \ni & 
  + \fr12 \frac{m_0^2}{N_c} V \bar \Phi_0^2 
 ~~\sim~~  
 \rmO(m_0^2)
 \rmO(\bar \Phi_0^2) 
 \rmO(\epsilon^{-2})\;, \la{qP1} \\ 
 & &  - i\, m K  V\, \bar \Phi_0 \tr \Bigl[ 
  e^{i\theta/\Nf} U_0 - U_0^\dagger e^{-i\theta/\Nf} \Bigr]
 ~~\sim~~ 
 \rmO(\bar K)
 \rmO(\bar \Phi_0)
 \rmO(\epsilon^{3}) \;.
 \la{quadPhi} \hspace*{0.5cm}
\ea 
Moreover, we want to carry out the computation at a fixed topology;
performing the integral over $\theta$ with the weight 
$\exp(i \theta \nu)$
introduces (after a shift) effectively one more term, 
\be
 \int \! {\rm d}^4 x\, \mathcal{L}_\rmi{$\chi$PT}  ~~\ni~~ 
 \frac{2 i \nu}{F}\bar
 \Phi_0 ~~\sim~~ \rmO(\bar \Phi_0) \rmO(\epsilon)
 \;.
 \la{linPhi}
\ee
Once the integral over $\theta$ is converted to a Gaussian  
over $\bar \Phi_0$, \eqs\nr{qP1}, \nr{linPhi} tell that the saddle
point is at leading order in $\epsilon$ at 
$\bar \Phi_0 \sim N_c \nu /(m_0^2 F V)\sim 
\rmO(\epsilon^3)/\rmO(m_0^2)$. Thus, 
we have fixed also the counting of $\bar \Phi_0$.

We can now collect together the full theory at fixed topology.
The factorized part of the zero-mode partition function becomes
\ba
 Z_\nu(\mu) & \propto & e^{-\frac{\nu^2}{2 \langle \nu^2 \rangle}}
 \int_\rmi{$U_0 \in$ U($\Nf$)} \hspace*{-0.5cm}
 {\det}^\nu U_0 \,
 \exp\Bigl[
 {\, \frac{\mu}{2} \tr (U_0+U_0^\dagger) + 2 \nu \frac{m K N_c}{m_0^2 F}
 \tr(U_0 - U_0^\dagger)
 }\Bigr] \la{Znu} \;, \hspace*{0.5cm}
\ea 
where
\be
 \frac{\langle \nu^2 \rangle}{V} = \frac{m_0^2 F^2}{4 N_c}
 \;. \la{onu2}
\ee
The first term in the exponent is $\rmO(1)$, 
while the latter is, as follows from~\eq\nr{quadPhi} with 
the given estimate of $\bar \Phi_0$, 
$\sim \rmO(\bar K)
 \rmO(\epsilon^6)/
 \rmO(m_0^2)$.
The non-zero momentum modes, on the other hand, are described by
\ba
  \int \! {\rm d}^4x\, \mathcal{L}_\rmi{$\chi$PT}	
  & \ni & + \int \! {\rm d}^4x\, 
 \frac{F^2}{4} \tr \Bigl[ 
 \partial_\mu U_\xi \partial_\mu U_\xi^\dagger
 \Bigr] 
 ~~ \sim ~~  
 \rmO(\epsilon^0)\Bigl[ 1+
 \rmO(\epsilon^4)\Bigr]\;, \la{kin} \\
 & & - \int \! {\rm d}^4x\, 
 \frac{m\Sigma}{2}
 \left. 
 \tr \Bigl[ 
 U_\xi U_0 + U_0^\dagger U_\xi^\dagger
 \Bigr]
 \right|_{\rmO(\xi^2)} 
 ~~ \sim ~~  
 \rmO(\epsilon^4)\;, \\
 & & + \int \! {\rm d}^4x\, 
 \frac{\alpha}{2 N_c} \Bigl(\partial_\mu \tilde \Phi_0 \Bigr)^2 
 ~~ \sim ~~  
 \rmO(\alpha)
 \rmO(\epsilon^2)\;, \\ 
 & & + \int \! {\rm d}^4x\,
 \frac{m_0^2}{2 N_c} \tilde \Phi_0^2  
 ~~ \sim ~~  
 \rmO(m_0^2)
 \rmO(\epsilon^0) \;. \la{m0}
\ea

To finalize the setup, one has to decide what kind
of counting rules are chosen for the parameters $m_0^2,\alpha,\bar K$. 
For simplicity, we will assume that $\rmO(\alpha)\sim 
\rmO(\bar K)\sim 1$. The counting of $m_0^2$
then leads to three distinct possibilities: 
\begin{itemize}
\item[1.]
In the unquenched theory, $m_0^2$ can be taken as ``large'', say 
$m_0^2 \sim \epsilon^{-2}$. Parametrically, then, 
$m_0^2/N_c \gg p^2$. In this case the term in~\eq\nr{m0} dominates
the action: 
the non-zero modes $\tilde \Phi_0$ (representing the $\eta'$)
are heavy and can be integrated 
out, resulting in the simple usual chiral theory following from~\eq\nr{XPT}. 

This choice is not available in the quenched limit, however: 
the field $\tilde \Phi_0$ cannot be integrated out~\cite{BG,S}, 
but {\em has} to be treated as a light degree of freedom.
Therefore, it is convenient to assign a different counting to it.  

\item[2.]
In the ``standard'' version of quenched chiral perturbation theory, 
one chooses $m_0^2 \sim \rmO(1)$, such that $m_0^2/N_c \sim
p^2$~\cite{ddhj}. Then \eq\nr{m0} is of the same order as the usual
kinetic terms following from~\eq\nr{kin}.
The term with $K$ in~\eq\nr{Znu}, on the other hand, 
can be neglected, since it is $\rmO(\epsilon^6)$
with this counting. 

This standard counting suffers from some problems, 
however. First of all, the integrals over the graded group of the 
supersymmetric formulation do not appear to be, strictly speaking, 
well defined~\cite{ddhj}, because the masses related to quadratic 
fluctuations, treated according to Zirnbauer's prescription~\cite{Z}, 
are not positive-definite (cf.\ \eq(3.8) in~\cite{ddhj}). 
Second,  Q$\chi$PT leads to a perturbative
expansion parameter $\sim m_0^2 /(N_c p^2)$. With the standard
counting this is of order unity, formally spoiling the convergence. 

\item[3.]
Because of the problems of the standard counting, we will 
consider an ``alternative'' counting here. In the alternative counting, 
$m_0^2$ is treated as a small quantity, say $m_0^2 \sim \epsilon^2$. 
Then~\eq\nr{m0} is formally a perturbation and the leading order
quadratic form {\em is} well defined.
This counting makes also explicit the fact that Q$\chi$PT
should only work in the window of~\eq\nr{Qwindow}.
With the choice $m_0^2 \sim \epsilon^2$, contributions from  
the coefficient $K$ should be kept in the results.

\end{itemize}
In the real world, obviously, $1/N_c$ is not tunable, and not
necessarily small. Therefore, the success or failure of the frameworks
described remains ultimately to be judged empirically, by comparing 
them with data. 
The expressions below are, for generality, 
for the ``standard counting'', while in the actual text we only
showed truncated versions, where terms of higher order 
according to the ``alternative counting'' had been dropped.

%
\section{Detailed results for full chiral perturbation theory}
\la{app:xpt}

\subsection{Space-time integrals appearing}
\la{app:ints}

After integration over the spatial volume, the time dependence of~\eq\nr{Cx}
appears in the forms
\ba 
 \int \! {\rm d}^3 \vec{x} \,
 & = & L^3  
 \;, \la{i1} \\ 
 \int \! {\rm d}^3 \vec{x} \,
 G(x) & = & T h_1(\tau)
 \;, \\ 
 \int \! {\rm d}^3 \vec{x} \,
 \Bigl[ G(x) \Bigr]^2
 & = & \frac{T^2}{L^3} g_1(\tau)
 \;, \\ 
 \int \! {\rm d}^3 \vec{x} \, \int \! {\rm d}^4 y \, 
 G(x-y) G(y) & = & 
 - T^3 h_2(\tau) 
 \;, \la{i4}
\ea
where
\ba
 h_1(\tau) & \equiv & \frac{1}{2}  
 \left[\left(\tau - {1 \over 2}\right)^2 - {1 \over 12}\right]
 \;, \la{h1} \\
 h_2(\tau) & \equiv & \frac{1}{24} 
 \left[\tau^2 \left(\tau - 1\right)^2 - {1 \over 30}\right] 
 \;, \la{h2} \\
 g_1(\tau) & \equiv & [h_1(\tau)]^2 + 
 \sum_{{\vec n} \neq 0} 
 \left[ {{\cosh}\left( |{\vec p}| (\tau -1/2) \right) \over 2 |{\vec p}| 
 {\sinh}\left(|{\vec p}|/2\right) } \right]^2 
 \;. \la{g1}
\ea
Here 
\ba
 |{\vec p}| = 2 \pi \frac{T}{L} \left[ \sum_{i=1}^3 
 n_i^2 \right]^{1/2}
 \;. \la{defp}
\ea

\subsection{Zero-mode integrals appearing}
\la{zm_full} \la{app:zeromode}

The zero-momentum mode integrals at fixed topology are related to the 
partition function
\be 
 Z_\nu(\mu) \equiv 
 \int_{U_0 \in {\rm U}(\Nf)}
 {\det}^\nu U_0 \, e^{\,\mu \re \tr U_0}
 \;, \la{Z0nu} 
\ee
where $\mu = m \Sigma V$.
The value of $Z_\nu(\mu)$ is known~\cite{brower,ls} to be
\be
 Z_\nu(\mu) = \det[I_{\nu+j-i}(\mu)],  
\ee 
where the determinant is taken over an $\Nf \times \Nf$ matrix, whose 
matrix element $(i,j)$ is the modified Bessel function $I_{\nu+j-i}$.
We will express our results in terms of the derivatives of this 
partition function, in particular
\be
 \sigma_\nu(\mu) \equiv \frac{\Sigma_\nu(\mu)}{\Sigma} \equiv 
 \frac{1}{\Nf} 
 \frac{\partial}{\partial \mu} \ln Z_\nu(\mu)
 \;. 
\ee
At small $\mu$ and non-zero $\nu$, 
\be
 \sigma_\nu(\mu) \approx \frac{|\nu|}{\mu} 
 \;.
 \la{Cnu}
\ee
Expectation values are denoted by
\be
 \langle ... \rangle^\mu_\nu \equiv
 \frac{ \int_{U_0 \in {\rm U}(\Nf)} (...) 
 \det^\nu \! U_0 \, e^{\,\mu \re \tr U_0} }{ \int_{U_0 \in {\rm U}(\Nf)}
 \det^\nu \! U_0 \, e^{\,\mu \re \tr U_0} }
 \; ;
\ee
both the superscript and subscript 
in $\langle...\rangle^\mu_\nu$ are often left out. 

Given these definitions, all the emerging expectation 
values can be computed analytically, 
using the techniques discussed in Appendix~B of~\cite{ddhj}. The
small-$\mu$ (small-$m$) limits are then obtained by using~\eq\nr{Cnu}. We give
here a complete collection of the integrals appearing, up 
to third order in the matrices $U_0, U_0^\dagger$. 
The  expectation
values for complex-conjugated operators
are obtained from those shown simply by $\nu \to -\nu$. 
For the
small-$\mu$ limits we only show the values of order $1/\mu^n$, for
$n$ powers of $U_0, U_0^\dagger$:
\ba
 \langle \tr U_0 \rangle & = & 
 \Nf \Bigl[ 
 \sigma_\nu - \frac{\nu}{\mu}
 \Bigr] 
 \la{za} \\
 & \approx & 
 \frac{\Nf}{\mu}(|\nu| - \nu)
 \;, \la{zamu}  \\
 \langle \tr (U_0^2) \rangle & = & 
 \Nf \Bigl[
 1 - \frac{2(\Nf+\nu)}{\mu}
 \Bigl( 
 \sigma_\nu - \frac{\nu}{\mu}
 \Bigr) 
 \Bigr]
 \la{zc} \\ 
 & \approx & 
 -\frac{2 \Nf}{\mu^2}(\Nf + \nu)(|\nu|-\nu)
 \;, \la{zcmu} \\
 \langle 
 \tr U_0 \tr U_0^\dagger
 \rangle & = & 
 \Nf \Bigl[
 \sigma_\nu' + \Nf \sigma_\nu^2 + \frac{\sigma_\nu}{\mu} - 
 \Nf \frac{\nu^2}{\mu^2} 
 \Bigr] 
 \la{ze} \\ 
 & \approx & 
 0 \times \frac{\Nf}{\mu^2}
 \;, \la{zemu} \\
 \langle (\tr U_0)^2 \rangle & = & 
 \Nf \Bigl[ 
 \sigma_\nu' + \Nf \sigma_\nu^2 - (1 + 2 \Nf \nu)
 \frac{\sigma_\nu}{\mu}  + (2 + \Nf \nu) \frac{\nu}{\mu^2}
 \Bigr]
 \la{zf} \\ 
 & \approx & 
 - \frac{2 \Nf}{\mu^2}(1 + \Nf \nu)(|\nu|-\nu)
 \;, \la{zfmu} \\
 \langle \tr (U_0^3) \rangle & = & 
 \Nf \biggl\{ 
 -\frac{2\nu}{\mu^3} (2 \nu^2 + 5 \Nf \nu + 2 \Nf^2 + 2 ) 
 - \frac{1}{\mu} (2 \Nf + 3 \nu) \nn 
 & & + \Bigl[ 
 1 + \frac{2}{\mu^2} (2 \nu^2 + 6 \Nf \nu + 2 \Nf^2 + 1)
 \Bigr] \sigma_\nu 
 - \frac{2}{\mu} (\sigma_\nu' + \Nf \sigma_\nu^2)
 \biggr\}
 \la{zh} \\ 
 & \approx & 
 \frac{4 \Nf}{\mu^3}(1+\Nf^2 + 3 \Nf\nu + \nu^2)(|\nu|-\nu)
 \;, \la{zhmu} \\
 \langle \tr U_0 \tr (U_0^2) \rangle & = & 
 \Nf \biggl\{ 
 -\frac{2\nu}{\mu^3}
 \Bigl[(4+\Nf^2)\nu + \Nf(4+\nu^2)\Bigr] - \frac{2 + \Nf\nu}{\mu} \nn
 & & + 
 \Bigl[ 
 \Nf + \frac{2}{\mu^2} \Bigl( (3 + 2 \Nf^2)\nu + \Nf (3 + 2 \nu^2) \Bigr)
 \Bigr] \sigma_\nu 
 \nn & & 
 - 2 \frac{\Nf + \nu}{\mu}
 (\sigma_\nu' + \Nf \sigma_\nu^2)
 \biggr\} \la{zj} \\ 
 & \approx & 
 \frac{4 \Nf}{\mu^3}
 \Bigl[ 2 \Nf + (2 + \Nf^2)\nu + \Nf \nu^2\Bigr] (|\nu|-\nu)
 \;, \la{zjmu} \\
 \langle \tr U_0^\dagger \tr (U_0^2) \rangle & = & 
 \Nf \biggl\{
 \frac{2 \Nf \nu^2}{\mu^3} (\Nf + \nu) + 
 \frac{2 + \Nf \nu}{\mu} \nn
 & & 
 + \Bigl[ 
 \Nf - \frac{2}{\mu^2}(\Nf + \nu)
 \Bigr] \sigma_\nu 
 - \frac{2(\Nf + \nu)}{\mu}
 (\sigma_\nu' + \Nf \sigma_\nu^2)
 \biggr\} 
 \la{zk} \\ 
 & \approx & 
 0 \times \frac{\Nf}{\mu^3}
 \;, \la{zkmu} \\ 
 \langle (\tr U_0)^3 \rangle & = & 
 \Nf \Bigl[
 -\frac{\nu}{\mu^3} (8 + 6 \Nf \nu + \Nf^2 \nu^2) + \frac{3}{\mu^2}
 (1 + 3 \Nf \nu + \Nf^2 \nu^2 )\sigma_\nu \nn 
 & & 
 -\frac{3}{\mu} (1 + \Nf \nu) (\sigma_\nu' + \Nf \sigma_\nu^2)
 + \sigma_\nu'' + 3 \Nf \sigma_\nu' \sigma_\nu + \Nf^2 \sigma_\nu^3 
 \Bigr]
 \la{zn} \\
 & \approx & 
 \frac{4 \Nf}{\mu^3}
 (2 + 3 \Nf \nu + \Nf^2 \nu^2)(|\nu|- \nu)
 \;, \la{znmu} \\
 \langle (\tr U_0^\dagger) (\tr U_0)^2 \rangle 
 & = & 
 \Nf \Bigl[
 \Nf \frac{\nu^2}{\mu^3} (2 + \Nf \nu) - \frac{1}{\mu^2}
 (1 + \Nf \nu + \Nf^2 \nu^2) \sigma_\nu \nn 
 & & 
 + \frac{1 - \Nf \nu}{\mu} (\sigma_\nu' + \Nf \sigma_\nu^2)
 + \sigma_\nu'' + 3 \Nf \sigma_\nu' \sigma_\nu + \Nf^2 \sigma_\nu^3  
 \Bigr] 
 \la{zo} \\ 
 & \approx & 
 0 \times \frac{\Nf}{\mu^3}
 \;. \la{zomu}
\ea
It may be noted from the small-$\mu$ expressions that plenty of
degeneracies emerge if we put $\Nf \to 1$: this is simply because
taking a trace has then no meaning.

\subsection{Results for the coefficients in \eq\nr{Cx}}
\la{app:coeffs}

Let us define 
\be
 \Sigma' \equiv \Sigma
 \biggl[
 1 - \frac{\Nf^2-1}{\Nf} \frac{G(0)}{F^2} 
 \biggr] = \Sigma
 \biggl[
 1 + \frac{\Nf^2-1}{\Nf} \frac{\beta_1}{F^2\sqrt{V}} 
 \biggr] \;,
 \la{primed}
\ee
and 
\be
 \mu' \equiv m \Sigma' V 
 \;. \la{primed2}
\ee
For the coefficient $C_I$ defined in~\eq\nr{Cx}, we then obtain (at NLO)
\be
 C_I = 
 -\fr14 \Bigl( \Sigma' \Bigr)^2
 \Bigl\langle
 \Bigl\{ \tr \Bigl[ 
 T^I(U_0 - U_0^\dagger)
 \Bigr] \Bigr\}^2
 \Bigr\rangle^{\mu'}_\nu 
 \;,
 \la{CI} 
\ee
where $I$ is not
summed over. For $I=0$ (flavour singlet), 
the result is immediately related 
to the expectation values listed in~\eqs\nr{za}--\nr{zomu}; for $I=a$,
one can make the connection by using independence of $a$ and the 
completeness relation 
\be
 \langle T^a_{ij} T^{a}_{kl} F_{ijkl}\rangle
 = \frac{1}{\Nf^2-1} \sum_{a=1}^{\Nf^2 - 1}
 \langle T^a_{ij} T^{a}_{kl} F_{ijkl}\rangle
 = \frac{1}{2(\Nf^2-1)} 
 \Bigl\langle F_{ijji} - \frac{1}{\Nf} F_{iijj} \Bigr\rangle 
 \;.
\ee 
We show explicitly only the small-$\mu$ limits here:
\ba
 C_0 & = & \frac{\Sigma^2 \Nf |\nu|}{\mu^2} 
 (1 - \Nf |\nu|)
 \;, \la{C0} \\
 C_a & = & \frac{\Sigma^2|\nu|}{2 \mu^2} 
 \;. \la{Ca}
\ea
For $\alpha_I$, we obtain, in a similar way, 
\ba
 \alpha_I & = & 
 \frac{\Sigma^2}{2 F^2}
 \biggl[ 
 1 - \frac{\Nf^2-2}{\Nf}\frac{G(0)}{F^2}
 \biggr]
 \times \nn 
 & & \times
 \biggl\langle
 \tr\Bigl[ 
 (U_0 T^I + T^I U_0^\dagger)^2 
 \Bigr]
 -\frac{1}{\Nf}
 \Bigl\{ \tr\Bigl[ 
 T^I (U_0 + U_0^\dagger)
 \Bigr] \Bigr\}^2
 \biggr\rangle^{\mu'}_\nu 
 \;. \la{alphaI}
\ea
For small $\mu$, 
\ba
 \alpha_0 & = & 
 \frac{2 \Sigma^2 |\nu|}{F^2\mu^2}
 \biggl[ 
 1 + \Nf 
 \frac{G(0)}{F^2}
 \biggr]
 (1-\Nf^2) 
 \;, \la{alpha0} \\
 \alpha_a & = & 
 \frac{\Sigma^2|\nu|}{F^2\mu^2 \Nf}
 \biggl[ 
 1 + \Nf 
 \frac{G(0)}{F^2}
 \biggr]
 (1 + \Nf |\nu|) 
 \;. \la{alphaa} 
\ea
For $\beta_I$, we obtain
\be
 \beta_I = 
 - \frac{\Sigma^2}{F^4}\biggl\langle
 \frac{\Nf^2+2}{4\Nf^2}
 \Bigl\{ \tr \Bigl[ T^I  (U_0 - U_0^\dagger ) \Bigr] \Bigr\}^2
 + \frac{\Nf^2-4}{4 \Nf}
 \tr \Bigl[ 
 (U_0 T^I - T^I U_0^\dagger)^2
 \Bigr] 
 \biggr  \rangle^\mu_\nu
 \;. \la{betaI}
\ee
For small $\mu$, 
\ba
 \beta_0 & = & 
 \frac{\Sigma^2|\nu|}{F^4\mu^2 \Nf}
 (\Nf^2-1)
 (\Nf^2 - 2 - 2 \Nf |\nu|)
 \;, \la{beta0} \\
 \beta_a & = & 
 \frac{\Sigma^2|\nu|}{2 F^4\mu^2 \Nf^2}
 \Bigl[
 \Nf^2 + 2 - \Nf (\Nf^2-4) |\nu|
 \Bigr] 
 \;. \la{betaa}
\ea
Finally, $\gamma_I$ reads
\ba
 \gamma_I & = & \frac{\Sigma^2}{2 F^4 V}
 \biggl\langle 
 \Bigl\{ \tr \Bigl[ T^I (U_0 + U_0^\dagger) \Bigr] \Bigr\}^2
 - \Nf 
  \tr \Bigl[(U_0 T^I + T^I U_0^\dagger)^2 \Bigr]
 \nn & & 
 -\mu \tr \Bigl[  (U_0 + U_0^\dagger) (U_0 T^I + T^I U_0^\dagger)^2
  \Bigr] 
 \nn & & 
 + \frac{2\mu}{\Nf}
 \tr \Bigl[ (U_0 + U_0^\dagger) (U_0 T^I + T^I U_0^\dagger) 
  \Bigr]
 \tr \Bigl[ T^I (U_0 + U_0^\dagger) \Bigr]
 \nn & & 
 - \frac{\mu}{\Nf^2}
 \tr (U_0 + U_0^\dagger )
 \Bigl\{ \tr \Bigl[ T^I (U_0 + U_0^\dagger) \Bigr] \Bigr\}^2
 \biggr\rangle^\mu_\nu
 \;. \la{gammaI}
\ea
For small $\mu$, 
\ba
 \gamma_0 & = & - \frac{2 \Sigma^2|\nu|}{F^4 V\mu^2 \Nf}
 (\Nf^2-1)
 (\Nf^2-4-2 \Nf |\nu| )
 \;, \la{gamma0} \\
 \gamma_a & = & \frac{\Sigma^2|\nu|}{F^4 V\mu^2 \Nf^2}
 \Bigl[
 \Nf^2(1-2\nu^2) - 4 + \Nf(\Nf^2-6)|\nu|
 \Bigr] 
 \;. \la{gammaa}
\ea

\section{Detailed results for quenched chiral perturbation theory}
\la{app:qxpt}

\subsection{Additional space-time integrals in the quenched theory}
\la{app:qints}

Apart from the integrals in~\eqs\nr{i1}--\nr{i4}, \eqs\nr{qCx} and \nr{qprop} 
imply that in the quenched case we need, in general,  
the following further ones:
\ba
 \int \! {\rm d}^3 \vec{x} \,
 F(x) & = & -T^3 h_2(\tau)
 \;, \\ 
 \int \! {\rm d}^3 \vec{x} \,
 G(x) F(x)
 & = & \frac{T^4}{L^3} g_2(\tau)
 \;, \\ 
 \int \! {\rm d}^3 \vec{x} \,
 \Bigl[ F(x) \Bigr]^2
 & = & \frac{T^6}{L^3} g_3(\tau)
 \;, \\ 
 \int \! {\rm d}^3 \vec{x}  \int \! {\rm d}^4 y \, 
 G(x-y) F(y) & = & 
 T^5 h_3(\tau)
 \;, \\
 \int \! {\rm d}^3 \vec{x}  \int \! {\rm d}^4 y \, 
 F(x-y) F(y) & = & 
 - T^7 h_4(\tau) 
 \;,
\ea
where we have defined (following the notation in~\eqs\nr{h1}--\nr{defp}), 
\ba
 h_3(\tau) & \equiv & \frac{1}{720} 
 \left[ \tau^2 \left(\tau - 1\right)^2 
 \left(\tau (\tau - 1) - {1 \over 2}\right) + {1 \over 42}\right] 
 \;, \\
 h_4(\tau) & \equiv & \frac{1}{120960} 
 \left[ \tau^2 \left(\tau - 1\right)^2 
 \left(3 \tau^4 - 6 \tau^3 - \tau^2 + 4 \tau + 2\right) - {1 \over 10}\right]
 \;, \la{h4} \\
 g_2(\tau) & \equiv & -h_1(\tau) h_2(\tau) 
 - \sum_{{\vec n} \neq 0} 
 {{\cosh}\left( |{\vec p}| (\tau -1/2) \right) \over 2 |{\vec p}| 
 {\sinh}\left(|{\vec p}|/2\right) } 
 {1\over 2 |{\vec p}|} 
 {{\rm d} \over {\rm d}|{\vec p}|} 
 \left({{\cosh}\left( |{\vec p}| (\tau -1/2) \right) \over 2 |{\vec p}| 
 {\sinh}\left(|{\vec p}|/2\right) }\right) 
 \;, \hspace*{0.5cm} \la{g2} \\
 g_3(\tau) & \equiv & [h_2(\tau)]^2 
 + \sum_{{\vec n} \neq 0} 
 \left[ {1\over 2 |{\vec p}|} 
 {{\rm d} \over {\rm d}|{\vec p}|} 
 \left({{\cosh}\left( |{\vec p}| (\tau -1/2) \right) \over 2 |{\vec p}| 
 {\sinh}\left(|{\vec p}|/2\right) }\right) \right]^2
 \;. \la{g3}
\ea

\subsection{Quenched zero-mode integrals for the flavour singlets}
\la{app:qzeromode}

As discussed in the main text, in the quenched case,
the results for the flavour non-singlet 
follow from those for the flavour singlet. Therefore, 
we only need to address the zero-mode
integrals arising for the flavour singlets, and do not 
present a similarly exhaustive list as in~Appendix~\ref{zm_full}.

The flavour singlets contain two parts, a connected contraction
($\propto\Nv$) and a disconnected one ($\propto \Nv^2$). 
The replica trick (cf.\ \cite{currents}) allows to obtain the result
for the connected contraction from a certain limit of U($\Nf$) integrals, 
discussed in~Appendix~\ref{zm_full}. For the disconnected contraction, 
on the other hand, the zero-mode integrals have to be honestly carried
out, for $\Nv = 1$, using the supersymmetric formulation of Q$\chi$PT. 
(The non-zero momentum modes of the Goldstone bosons can 
still be treated with the replica formulation~\cite{ds}, 
and only the remaining 
zero-momentum mode integrals need to be transformed to the supersymmetric 
ones.) We first list the supersymmetric integrals for $\Nv = 1$, 
and then the generalizations to any $\Nv$ obtained with the replica trick.

Let us start with some notation. We introduce 
a projection operator $\Pv$, 
\be
 (\Pv)_{ij} \equiv \left\{
 \begin{array}{ll}
 \delta_{ij}, & \mbox{for $i,j$ = physical flavours in the valence block,} \\
 0          , & \mbox{otherwise.}
 \end{array} 
 \right. 
\ee
Using again the scaling variable $\mu\equiv m\Sigma V$, 
all mass dependence of the results can be expressed 
in terms the same zero-mode integral as appears in the quark 
condensate obtained with $\widehat{\rm Gl}(1|1)$ \cite{dotv}: 
\be
 \frac{1}{2\Nv} \langle \str[\Pv ( U_0 + U_0^{-1} )] \rangle \equiv
 \sigma_\nu \equiv 
 {\Sigma_\nu (\mu) \over \Sigma}  = 
 \mu \Bigl[ I_\nu (\mu) K_\nu (\mu) + I_{\nu +1} (\mu)
 K_{\nu -1}(\mu) \Bigr] +{\nu \over \mu}
 \;, 
 \la{zerocon} 
\ee
where $I_\nu, K_\nu$ are modified
Bessel functions. We recall that, for $\nu\neq 0$,
\be
 \sigma_\nu(\mu) \approx \frac{|\nu|}{\mu} 
 \;,
 \la{smallmu} \la{Cqnu}
\ee
as in~\eq\nr{Cnu}. 
Note that, in contrast to~\eq\nr{zerocon}, 
\be
 \langle \str ( U_0 + U_0^{-1} ) \rangle = 0 \;,
 \la{fullcond}
\ee
and also that $\langle \str ( U_0 - U_0^{-1} ) \rangle = 0$.

The zero-mode integrals for
$\Nv = 1$ can be derived following the techniques
discussed in~\cite{ddhj}, particularly the explicit
parametrization of $\widehat{\rm Gl}(1|1)$.  
The integrals needed, and their small-$\mu$
limits, read ($U\equiv U_0$ here)
\ba
 \Bigl\langle
 (U_{11})^2+(U^{-1}_{11})^2 
 \Bigr\rangle & = & 
 2 \Bigl[ 
 \sigma_\nu' - \frac{\sigma_\nu}{\mu} + 1 + \frac{2\nu^2}{\mu^2}
 \Bigr]
 \\ 
 & \approx & 
 \frac{4|\nu|}{\mu^2} (|\nu| - 1)
 \;, \\
 \Bigl\langle
 (U_{11})^2-(U^{-1}_{11})^2 
 \Bigr\rangle & = & 
 \frac{4\nu}{\mu} \Bigl[ 
 \frac{1}{\mu} - \sigma_\nu
 \Bigr]
 \\ 
 & \approx & 
 \frac{4\nu}{\mu^2} (1 - |\nu|)
 \;, \\
 \Bigl\langle
 U_{11} U^{-1}_{11}
 \Bigr\rangle & = & 
 \sigma_\nu' + \frac{\sigma_\nu}{\mu} + 1
 \\ 
 & \approx & 
 0 \times \frac{1}{\mu^2}
 \;, \\
 \Bigl\langle
 (U_{11})^3 + (U^{-1}_{11})^3  
 \Bigr\rangle & = & 
 \sigma_\nu''-7 \frac{\sigma_\nu'}{\mu} + \sigma_\nu
 \Bigl[
 2 + \frac{7 + 8 \nu^2}{\mu^2} 
 \Bigr]- \frac{6}{\mu} - \frac{24\nu^2}{\mu^3} \hspace*{0.8cm}
 \\ 
 & \approx & 
 \frac{8|\nu|}{\mu^3}(|\nu|-1)(|\nu|-2)
 \;, \\
 \Bigl\langle
 (U_{11})^2 U^{-1}_{11}
 + U_{11} (U^{-1}_{11})^2
 \Bigr\rangle & = & 
 \sigma_\nu'' + \frac{\sigma_\nu'}{\mu}
 + \sigma_\nu \Bigl[ 2 - \frac{1}{\mu^2} \Bigr] + \frac{2}{\mu}
 \\ 
 & \approx & 
 0 \times \frac{1}{\mu^3}
 \;, \\
 \Bigl\langle
 (U_{11} + U^{-1}_{11})
 (U_{12}U_{21} + U^{-1}_{12} U^{-1}_{21})
 \Bigr\rangle & = & 
 2 \Bigl[ 
 -\sigma_\nu''+\frac{\sigma_\nu'}{\mu}-\frac{\sigma_\nu}{\mu^2}+\frac{2}{\mu}+
 \frac{4\nu^2}{\mu^3}
 \Bigr]
 \\ 
 & \approx & 
 \frac{8|\nu|}{\mu^3}(|\nu|-1)
 \;, \\
 & & \hspace*{-5cm}
 \Bigl\langle
 \Bigl[
 U_{11} + U^{-1}_{22} - U^{-1}_{11}  - U_{22} 
 \Bigr]
 \Bigl[ 
 (U_{11})^2 + (U^{-1}_{11})^2
 \Bigr]
 \Bigr\rangle = 
 \nn 
 & = & \frac{16 \nu}{\mu^2}
 \Bigl[ \sigma_\nu - \frac{1}{\mu}
 \Bigr]
 \\
 & \approx & 
 \frac{16\nu}{\mu^3}(|\nu| - 1)
 \;. \la{Fthree}
\ea

Using these integrals together with the $\Nf\to 0$ limits of
the corresponding U($\Nf$) integrals from Appendix~\ref{app:zeromode} 
(obtained, in each case, with the replacements $\str\to \tr$, 
$U_0^{-1} \to U_0^\dagger$, $\Pv \to 1$), 
we can deduce that
\ba
 \Bigl\langle \str \Bigl[ 
 ( U_0 \Pv + \Pv U_0^{-1})^2 
 \Bigr] \Bigr\rangle & = & 
 4 \Nv \biggl[ 
 1 + \frac{\nu^2}{\mu^2} + 
 \frac{\Nv}{2}
 \biggl(
 \sigma_\nu' - \frac{\sigma_\nu}{\mu} 
 \biggr)
 \biggr]
 \la{z1} \\
 & \approx & 
 \frac{4\Nv|\nu|}{\mu^2}
 \Bigl[ 
 |\nu| - \Nv
 \Bigr]
 \;, \la{z1mu} \\
 \Bigl\langle \str \Bigl[ 
 ( U_0 \Pv - \Pv U_0^{-1})^2 
 \Bigr] \Bigr \rangle & = & 
 4 \Nv \biggl[ 
 \frac{\nu^2}{\mu^2} + \frac{\Nv}{2}
 \biggl( 
 \sigma_\nu' - \frac{\sigma_\nu}{\mu} 
 \biggr)
 \biggr]
 \la{z3} \\
 & \approx &
 \frac{4 \Nv |\nu|}{\mu^2}
 \Bigl[ 
 |\nu| - \Nv
 \Bigr] 
 \;, \la{z3mu} \\
 \Bigl\langle \Bigl[\str \Pv (U_0 + U_0^{-1} ) \Bigr]^2 \Bigr\rangle
 & = & 4 \Nv \Bigl[\sigma_\nu' 
       + \Nv \Bigl( 1 + \frac{\nu^2}{\mu^2}\Bigr) \Bigr]
 \la{z2} \\
 & \approx &
 \frac{4 \Nv |\nu|}{\mu^2}
 \Bigl[ 
 -1 + \Nv |\nu|
 \Bigr] 
 \;, \la{z2mu} \\
 \Bigl \langle \Bigl[\str \Pv (U_0 - U_0^{-1} ) \Bigr]^2 \Bigr\rangle
 & = & 4 \Nv \Bigl[  \Nv \frac{\nu^2}{\mu^2} -\frac{\sigma_\nu}{\mu} 
       \Bigr] 
 \la{z4} \\
 & \approx & 
 \frac{4 \Nv |\nu|}{\mu^2}
 \Bigl[ 
 -1 + \Nv |\nu|
 \Bigr] 
 \;, \la{z4mu} \\
  & & \hspace*{-5cm} \Bigl \langle \str \Bigl[ (U_0 \Pv + \Pv U_0^{-1} )
 (U_0 \Pv - \Pv U_0^{-1} ) \Bigr] \Bigr\rangle =  
 \nn & = &
 \frac{4 \Nv \nu}{\mu}
 \Bigl[\frac{\Nv}{\mu} - \sigma_\nu \Bigr] \\
 & \approx & 
 \frac{4 \Nv \nu}{\mu^2}
 \Bigl[ \Nv - |\nu| \Bigr]
 \;, \\
  & & \hspace*{-5cm} \Bigl \langle  
 \Bigl[ \str \Pv (U_0 + U_0^{-1} ) \Bigr]
 \Bigl[ \str \Pv (U_0 - U_0^{-1} ) \Bigr] \Bigr\rangle =  
 \nn & = &
 \frac{4 \Nv \nu}{\mu}
 \Bigl[\frac{1}{\mu} - \Nv \sigma_\nu \Bigr] \\
 & \approx & 
 \frac{4 \Nv \nu}{\mu^2}
 \Bigl[ 1 - \Nv |\nu| \Bigr]
 \;, \\
 & & \hspace*{-5cm} \Bigl\langle
 \str\Bigl\{ 
 (U_0+U_0^{-1}) 
 (U_0 \Pv + \Pv U_0^{-1})^2
 \Bigr\} 
 \Bigr\rangle = 
 \nn & = &
 4 \Nv\biggl[
 -\frac{\sigma_\nu'}{\mu} + 
 \sigma_\nu \Bigl( 2 + \frac{1 + 2 \nu^2}{\mu^2}
 \Bigr) 
 - \frac{\Nv}{\mu} \biggl( 1 + \frac{4\nu^2}{\mu^2} 
 \biggr)
 \biggr] 
 \la{z5} \\
 & \approx & 
 \frac{8 \Nv |\nu|}{\mu^3}
 \Bigl[
 1 + \nu^2 - 2 \Nv |\nu|
 \Bigr]
 \;, \la{z5mu} \\
 & & \hspace*{-5cm}
 \Bigl\langle
 \str\Bigl[ (U_0+U_0^{-1})
 (U_0 \Pv+\Pv U_0^{-1}) \Bigr]
 \str \Bigl[ \Pv
 (U_0+U_0^{-1}) \Bigr]
 \Bigr\rangle = 
 \nn 
 & = & 
 4 \Nv \biggl\{
 -\frac{4\nu^2}{\mu^3} + {\Nv} \biggl[
 - \frac{\sigma_\nu'}{\mu} 
 + \sigma_\nu\Bigl( 2 + \frac{1 + 2 \nu^2}{\mu^2}
 \Bigr) 
 \biggr] 
 \biggr\}
 \la{z6} \\
 & \approx & 
 \frac{8 \Nv |\nu|}{\mu^3}
 \Bigl[ 
 - 2 |\nu| + \Nv 
 \Bigl( 1 + \nu^2
 \Bigr)
 \Bigr]
 \;, \la{z6mu} \\
 & & \hspace*{-5cm} \Bigl\langle
 \str
 (U_0+U_0^{-1})
 \Bigl[ \str \Pv  
 (U_0+U_0^{-1}) \Bigr]^2 
 \Bigr\rangle = 
 \nn & = & 
 2 \frac{{\rm d}}{{\rm d}\mu}  
 \Bigl\langle \Bigl[\str \Pv (U_0 + U_0^{-1} ) \Bigr]^2 
 \Bigr\rangle
 \la{z9} \\
 & \approx & 
 \frac{16 \Nv |\nu|}{\mu^3}
 \Bigl[ 
 1 - \Nv |\nu|
 \Bigr] 
 \;, \la{z9mu} \\
& & \hspace*{-5cm} \Bigl\langle
 \str
 (U_0-U_0^{-1})
 \str \Bigl[ 
 (U_0 \Pv + \Pv U_0^{-1})^2 \Bigr]
 \Bigr\rangle =  
 \nn & = & 
 \frac{16 \Nv \nu}{\mu^2} \biggl[ 
 \sigma_\nu - \frac{\Nv}{\mu}
 \biggr]
 \la{zK} \\
 & \approx & 
 \frac{16\Nv \nu }{\mu^3}
 \Bigl[ 
 |\nu| - \Nv
 \Bigr]
 \;. \la{zKmu} 
\ea
In 
\eq\nr{z9}
we used \eq\nr{fullcond} 
together with the fact that, in general, 
\be
 \Bigl\langle \Bigl[ \str (U_0 + U_0^{-1} )\Bigr] M \Bigr\rangle 
 = 
 \Bigl\langle \str (U_0 + U_0^{-1} ) \Bigr\rangle \Bigl\langle M \Bigr\rangle 
 + 2 \frac{{\rm d}}{{\rm d}\mu} \langle M \rangle \;. 
\ee

\subsection{Results for the coefficients in \eq\nr{qCx}}
\la{app:qcoeffs}

Given these building blocks, we can collect our results together. 
In analogy with \eqs\nr{primed}, \nr{primed2}, we define
\be
 \Sigma' \equiv \Sigma
 \biggl[
 1 + \frac{E(0)}{F^2} 
 \biggr] = 
 \Sigma
 \biggl\{
 1 + \frac{1}{2 N_c F^2}
 \Bigl[
 \alpha G(0)  + m_0^2 F(0) 
 \Bigr]
 \biggr\}
 \;,
 \la{qprimed}
\ee
and 
\be
 \mu' \equiv m \Sigma' V 
 \;. \la{qprimed2}
\ee
Then the results for the coefficients in~\eq\nr{qCx},
together with the small-$\mu$ limits, read: 
\ba
 C_0 & = & 
 -\fr14 \Bigl( \Sigma' \Bigr)^2 
 \Bigl\langle 
 \Bigl[\str \Pv (U_0 - U_0^{-1} ) \Bigr]^2
 \Bigr\rangle^{\mu'}_\nu
 \la{qC0} \\ 
 & \approx & 
 \frac{\Sigma^2 \Nv |\nu|}{\mu^2} (1 - \Nv |\nu|)
 \;, \la{C0mu} \\
 \alpha_0 & = & 
 \frac{(\Sigma')^2}{2 F^2}
 \Bigl\langle \str\Bigl[ 
 ( U_0 \Pv + \Pv U_0^{-1})^2 
 \Bigr] \Bigr\rangle^{\mu'}_\nu 
 + \Bigl[\frac{K\Sigma}{F} - \frac{\Sigma^2}{2 F^4} G(0) \Bigr] 
 \Bigl
 \langle \Bigl[\str \Pv (U_0 + U_0^{-1} ) \Bigr]^2 
 \Bigr \rangle^\mu_\nu \nn
 & & 
 + \frac{4 K N_c \Sigma \nu}{m_0^2 F^3 V} 
 \biggl\{ \Bigl\langle \str
 \Bigl[ 
 (U_0 \Pv + \Pv U_0^{-1}) (U_0 \Pv - \Pv U_0^{-1})
 \Bigr]
 \Bigr\rangle^\mu_\nu 
 \nn
 & & \hphantom{\frac{4 K N_c \Sigma \nu}{m_0^2 F^3 V} 
 \biggl\{}
 + \frac{\mu}{4} 
 \Bigl\langle
 \str(U_0 - U_0^{-1})
 \str\Bigl[
 (U_0 \Pv + \Pv U_0^{-1})^2 
 \Bigr]
 \Bigr\rangle^\mu_\nu \biggr\} 
 \la{qalpha0} \\ 
 & \approx & 
 \frac{2 \Sigma^2 \Nv |\nu|}{F^2 \mu^2}
 \biggl\{
 |\nu| - \Nv 
 + \Bigl[ \frac{G(0)}{F^2} - \frac{2 F K}{\Sigma} \Bigr](1 - \Nv |\nu|) 
 \biggr\}
 \;, \la{alpha0mu} \\
 \alpha_0' & = & 
 -\frac{(\Sigma')^2}{2 F^2} 
 \Bigl
 \langle \Bigl[\str \Pv (U_0 + U_0^{-1} ) \Bigr]^2 
 \Big\rangle^{\mu'}_\nu
 \nn & &  
 -\frac{4 K N_c \Sigma \nu}{m_0^2 F^3 V} 
 \Bigl\langle
 \Bigl[ \str
  \Pv (U_0 + U_0^{-1}) 
 \Bigr]
 \Bigl[ \str
  \Pv (U_0 - U_0^{-1}) 
 \Bigr]
 \Bigr\rangle^\mu_\nu 
 \la{alpha0p} \\ 
 & \approx &
 \frac{2 \Sigma^2 \Nv |\nu|}{ F^2 \mu^2} \Bigl(1 - \Nv |\nu|\Bigr) 
 \biggl(
 1 - \frac{8 K N_c |\nu|}{m_0^2 F \Sigma V} 
 \biggr)
 \;, \la{alpha0pmu} \\
 \beta_0 & = & 
 -\frac{\Sigma^2}{4 F^4}
 \Bigl
 \langle \Bigl[\str \Pv (U_0 - U_0^{-1} ) \Bigr]^2 
 \Bigr\rangle^\mu_\nu
 \la{qbeta0} \\ 
 & \approx & 
 \frac{\Sigma^2 \Nv |\nu|}{F^4 \mu^2 } (1 - \Nv |\nu|) 
 \;, \la{beta0mu} \\
 \beta_0' & = & 
 \frac{\Sigma^2}{F^4}
 \Bigl \langle \str \Bigl[ 
 ( U_0 \Pv - \Pv U_0^{-1})^2 
 \Bigr] \Bigr \rangle^\mu_\nu 
 \la{beta0p} \\ 
 & \approx & 
 \frac{4 \Sigma^2 \Nv |\nu|}{F^4 \mu^2 } (|\nu| - \Nv ) 
 \;, \la{beta0pmu} \\
 \beta_0'' & = & 
 -\frac{\Sigma^2}{2 F^4}
 \Bigl \langle \Bigl[\str \Pv (U_0 - U_0^{-1} ) \Bigr]^2 
 \Bigr \rangle^\mu_\nu
 \la{beta0pp} \\
 & \approx & 
 \frac{2 \Sigma^2 \Nv |\nu|}{F^4 \mu^2 } (1 - \Nv |\nu|) 
 \;, \la{beta0ppmu} \\
 \gamma_0 & = & 
 -\frac{\Sigma^2}{2 F^4 V}
 \biggl\langle 
 -\fr13
 \Bigl[\str \Pv (U_0 + U_0^{-1} ) \Bigr]^2 
   \nn & & +  
  \mu\,
 \str\Bigl\{ 
 (U_0+U_0^{-1}) 
 (U_0  \Pv +  \Pv U_0^{-1})^2
 \Bigr\}
 \biggr\rangle^\mu_\nu 
 \la{qgamma0} \\
 & \approx & 
 - \frac{2 \Sigma^2 \Nv |\nu|}{F^4 V \mu^2} 
 \Bigl[ 
 \fr73 + 2 \nu^2 - \frac{13}{3} \Nv |\nu|
 \Bigr]
 \;, \la{gamma0mu} \\ 
 \gamma_0' & = &  
 \frac{\Sigma^2}{F^4 V}
 \, \mu \,
 \Bigl \langle
 \str\Bigl[ (U_0+U_0^{-1})
 (U_0 \Pv + \Pv U_0^{-1}) \Bigr]
 \str \Bigl[ \Pv
 (U_0+U_0^{-1}) \Bigr]
 \Bigr \rangle^\mu_\nu
 \la{gamma0p} \\
 & \approx & 
 \frac{8 \Sigma^2 \Nv |\nu|}{F^4 V \mu^2} 
 \Bigl[-2 |\nu| + \Nv
 \Bigl( 1 + \nu^2
 \Bigr)
 \Bigr] 
 \;, \la{gamma0pmu} \\
 \gamma_0'' & = & 
 -\frac{\Sigma^2}{F^4 V}
 \, \mu \frac{{\rm d}}{{\rm d}\mu}  
 \Bigl \langle \Bigl[\str \Pv (U_0 + U_0^{-1} ) \Bigr]^2 
 \Bigr \rangle^\mu_\nu
 \la{gamma0pp} \\
 & \approx & 
 - \frac{8 \Sigma^2 \Nv |\nu|}{ F^4 V \mu^2} 
 \Bigl[1 - \Nv |\nu|
 \Bigr] 
 \;. \la{gamma0ppmu} 
\ea


\end{document}